\documentclass[danish,a4paper,10pt,
twocolumn,
amsmath,amssymb,floatfix,superscriptaddress, 
aps,
pra,showpacs,longbibliography]{revtex4-1}

\usepackage{graphicx}
\usepackage[english]{babel}
\usepackage{bm}
\usepackage{braket}
\usepackage{float}
\usepackage{mathtools}
\usepackage{dcolumn}
\usepackage[draft]{fixme}
\usepackage{color}
\usepackage{lipsum}
\usepackage{tikz}
\usetikzlibrary{decorations.pathreplacing}
\definecolor{red}{rgb}{1,0,0}
\definecolor{blue}{rgb}{0,0,1}

\begin{document}

\title{Attosecond transient-absorption spectroscopy of polar molecules}
\author{J\o rgen~Johansen~R\o rstad}
\affiliation{Department of Physics and Astronomy, Aarhus University, 8000 Aarhus C, Denmark}
\author{Nikolaj~S.~W.~Ravn}
\affiliation{Department of Physics and Astronomy, Aarhus University, 8000 Aarhus C, Denmark}
\author{Lun~Yue}
\email{lun\_yue@msn.com}
\affiliation{Department of Physics and Astronomy, Louisiana State University, Baton Rouge, Louisiana 70803, USA}
\author{Lars~Bojer~Madsen}
\affiliation{Department of Physics and Astronomy, Aarhus University, 8000 Aarhus C, Denmark}

\date \today
\begin{abstract}
We apply attosecond transient absorption spectroscopy (ATAS) to explore the effects of a nonzero permanent dipole on electron dynamics at the subfemtosecond scale, exemplified in the polar LiF molecule. In contrast with nonpolar systems, a familiar feature of the ATA spectra---the light-induced structures---are observed adjacent to a bright state. Moreover, a previously unobserved ladder structure is identified. The new observations are analyzed in the context of a model based on fixed-nuclei adiabatic states, supported by full numerical simulations. Analytic calculations originating in the adiabatic model shed light on the nature and origins of the new findings. 
\end{abstract}

\maketitle

\section{Introduction}

In the advancing field of attosecond science \cite{bengtsson2017,calegari2016advances}, attosecond transient absorption spectroscopy (ATAS) has emerged as a potent technique for investigating electron motion on its natural time scale \cite{loh2008, goulielmakis2010, beck2015probing, wu2016theory}. In ATAS, two pulses are employed in conjunction; a femtosecond near-infrared (NIR) pulse and an attosecond ultraviolet (UV) pulse induce dynamics in a target system, and the accumulated signal of interference between the dipole response of the target and the incoming UV field is recorded---which forms the basis for further analysis. An important factor in this scheme is the relative phase of the two pulses, and it can be controlled with great precision by varying the temporal delay between them. Scanning over a range of delays enables the compilation of detailed spectrograms, exhibiting the emission and absorption of light. 

ATAS has been applied to increasingly complex systems, contributing to our understanding of several physical processes. Examples of applications in atoms include the observation of autoionization in Ar \cite{wang2010attosecond}, the tailoring of transient pulses \cite{wirth2011}, creation and manipulation of wave packets in He \cite{ott2014}, and others probings of subfemtosecond phenomena \cite{holler2011, pabst2012, kobayashi2017, sabbar2017}. In molecules, studies have been conducted on the effects of nuclear motion on bound electron dynamics \cite{baekhoej2015}, charge migration following ionization \cite{hollstein2017}, on systems containing conical intersections \cite{baekhoej2018}, and both theoretical and experimental investigations of the dynamics in H$_2$ \cite{cheng2016}, N$_2$ \cite{warrick2016, warrick2017}, and O$_2$ \cite{liao2017}. ATAS has also been successfully applied to solids \cite{schultze2014, borja2016, moulet2017} and to dense gases \cite{liao2015, liao2016}, which further emphasizes the general versatility of the technique. 

The spectrograms of ATAS are characterized by features which represent underlying physical processes in the system. The most common features have been studied extensively, and include light-induced structures (LISs) indicating the presence of virtual intermediate states involved in multi-photon processes \cite{chen2012, baekhoej2015_2}; oscillating fringes arising from the dressing of populated states by the NIR-pulse \cite{chen2013, chini2014}; hyperbolic sidebands adjacent to absorption lines, associated with perturbed free-induction decay \cite{lindberg1988, britocruz1988}; and modification of absorption lines \cite{santra2011, pfeiffer2012, chen2013lip, ott2013lorentz}, typically by Stark shifts \cite{chini2012}, such as Autler-Townes splitting \cite{autler1955, wu2013}. Our previous work \cite{rorstad2017} supplemented the existing studies with analytical expressions describing several of these features. 

In this paper we consider the application of ATAS to systems with a nonzero permanent dipole, exemplified by the polar heteronuclear molecule LiF. We show that the presence of a permanent dipole can lead to common features appearing in unconventional settings---while also giving rise to new features---in the ATA spectrograms. Our study sheds light on electron dynamics in an important class of systems possessing inherent permanent dipoles. We employ two approaches to simulate the dynamics of the system: One is based on the numerical propagation of the time-dependent Schr\"odinger equation (TDSE) including nuclear motion in a reduced basis of electronic Born-Oppenheimer states, whereas the other relies on a formalism of adiabatic states and assumes a frozen internuclear distance. From the latter we derive analytical and semi-analytical expressions that describe and explain the new findings. 

The paper is organized as follows. Section~\ref{section:theory} contains descriptions of all relevant models and methods. In Sec.~\ref{subsection:response} the response function is introduced; Sec.~\ref{subsec:TDSE} covers the method based on numerically solving the TDSE; Sec.~\ref{subsec:adiab} describes the derivation of the adiabatic model. In Sec.~\ref{sec:results} the results from the various models are presented. Specifically, Sec.~\ref{subsec:results:TDSE} features the results calculated by the TDSE-based numerical method; in Sec.~\ref{subsec:features} expressions describing the LISs and ladder structure are derived, and the resulting spectra displayed; and in Sec.~\ref{subsec:ovsa} we consider the differences in the spectra from an oriented and an aligned target. Section~\ref{section:conclusion} concludes the paper and gives an outlook. Appendices~\ref{section:appendix_UV} and \ref{section:appendix_IR} contain derivations of specific Fourier transforms related to the UV and NIR fields. Atomic units ($\hbar=e=m_e=a_0=1$) are used throughout, unless otherwise indicated.

\section{\label{section:theory}Theory}
This section starts with the introduction of the response function of the system, which is the origin of all ATA spectra in this paper. The TDSE-based numerical method is described, and the adiabatic model is presented. 

\subsection{\label{subsection:response}Single-system response function and field parameters}
Under the present convention, the NIR pulse center is fixed at time $t=0$, with the UV pulse centered at the relative time delay $t=\tau$, indicating that for positive (negative) delays the UV pulse will trail (precede) the NIR pulse. In the following we assume that the dynamics we are interested in can be adequately represented by a single-system model. This assumption implies that macroscopic propagation effects are not accounted for, an approximation which has been shown to be valid for sufficiently dilute gases \cite{gaarde2011}. In this case, the ATA spectrum is described by  
\begin{equation}
\tilde{S}(\omega,\tau) = \frac{4\pi \rho \omega}{c}\text{Im}[\tilde{F}^*_\text{UV}(\omega,\tau)\tilde{D}(\omega,\tau)] \label{theory:A:response_1},
\end{equation}
where $\rho$ is the density of the molecules in the target, which is arbitrarily set to unity, $\tilde{F}_\text{UV}(\omega,\tau)$ is the incoming UV pulse in the frequency domain, $c\simeq137$ is the speed of light, and $\tilde{D}(\omega,\tau)$ is the Fourier transform of the expectation value of the dipole moment's $z$-component, which coincides with the direction of the field polarization. The tilde denotes Fourier transformed quantities, with the definition $\tilde{f}(\omega)=\frac{1}{\sqrt{2\pi}}\int_{-\infty}^\infty dt\, f(t) e^{-i\omega t} $. A negative value of $\tilde{S}(\omega,\tau)$ implies absorption of light, a positive value implies emission. For a detailed derivation of Eq.~\eqref{theory:A:response_1} and how it relates to commonly used and equivalent expressions in the literature, we refer the reader to Refs.~\cite{baggesen2012} and \cite{baekhoej2015}, respectively. 

The two incoming electric fields are derived from 
\begin{equation}
A(t)=A_0\exp\left[-\frac{(t-t_c)^2}{T^2/4}\right]\cos{[\omega(t-t_c)+\varphi]}\label{theory:A:pulse},
\end{equation}
through the relation $F(t) = -\partial_tA(t)$. In Eq.~\eqref{theory:A:pulse} ${A_0=F_0/\omega}$, with $\omega$ the angular frequency and $F_0$ the amplitude, related to the field intensity $I$ through $I=|F_0|^2$; $t_c$ is the center of the pulse; $\varphi$ is the carrier-envelope phase; and $T=N_cT_c=N_c\frac{2\pi}{\omega}$ is the period of the pulse, with $N_c$ the number of cycles in the pulse and $T_c$ the period of a single cycle. $T$ is related to the full width at half maximum (FWHM) by $T_\text{FWHM}=\sqrt{\log{2}}T$. In the present work the following field parameters are used: $\lambda_\text{UV}=160$~nm, corresponding to $\omega_\text{UV}=7.75$~eV; $I_\text{UV}=5\times10^7$~W/cm${}^2$; $T_\text{UV}=1.07$~fs; $N_{c,\text{UV}}=2$; $\lambda_\text{NIR}=800$~nm, corresponding to $\omega_\text{NIR}=1.55$~eV; $I_\text{NIR}=10^{12}$~W/cm${}^2$; $T_\text{NIR}=40.01$~fs; $N_{c,\text{NIR}}=15$. Both fields are linearly polarized in the $z$ direction. The moderate intensities of the fields permit the use of certain weak-field approximations, which are described in detail where relevant.

\subsection{\label{subsec:TDSE}Full numerical model}
Here we consider a model of a diatomic polar molecule where the nuclear degrees of freedom are included via the internuclear separation, $R$, between the nuclei. Nuclear rotation is neglected due to its much longer time scale. The molecule is positioned such that its axis of symmetry coincides with the $z$-axis. 

In order to calculate the response function of the system [Eq.~\eqref{theory:A:response_1}], we must determine the expectation value of the dipole operator $\braket{\hat{D}}_{\text{tot}}(t) = \braket{\Psi(t) | \hat{D} | \Psi(t)}_{\text{tot}} $. Here $\Psi(t,\mathbf{r},R)$ is the total wave function, with $\mathbf{r}$ the electronic coordinates with respect to the center of mass of the nuclei, and $\hat{D}$ the operator representing the total dipole moment of the system. In this article we denote integration over both nuclear and electronic coordinates by $\braket{\cdot|\cdot}_{\text{tot}}$ and integration over the electronic coordinates only as $\braket{\cdot|\cdot}$.

The TDSE for the system after separating the center of mass motion is
\begin{equation}
	i\partial_t \Psi(t,\mathbf{r},R) = H(t) \Psi(t,\mathbf{r},R), \label{theory:B:FullTDSE}
\end{equation}
with the Hamiltonian
\begin{equation}
	H(t) = T_N + T_e + V_{NN} + V_{ee} + V_{Ne} + V_{I}(t) \label{theory:B:FullHamiltonian}.
\end{equation}
Here $T_N = - (1/M_N)\partial^2/\partial R^2$ is the kinetic energy of the nuclei with $M_N = M_1M_2/(M_1 + M_2)$ the reduced mass of the nuclei, $T_e$ the kinetic energies of the electrons, and $V_{NN}$, $V_{ee}$ and $V_{Ne}$ are the Coulomb potentials between the nuclei and electrons. Working in the length gauge and dipole approximation, the interaction with the field is $V_{I} = -F(t) \hat{D} $, where $F(t) = F_{\text{NIR}}(t) + F_{\text{UV}}(t) $ is the total field.

We expand the wave function in the lowest electronic Born-Oppenheimer states~\cite{yue2013,baekhoej2015}
\begin{equation}
	\Psi(t,\mathbf{r};R) = \chi_1(t;R)\psi_1(\mathbf{r};R) + \chi_2(t;R)\psi_2(\mathbf{r};R) \label{theory:B:NSurface}.
\end{equation}
This expansion is sufficient for the system, intensities, and frequencies of interest in the present work. Inserting the expansion of Eq.~\eqref{theory:B:NSurface} into the TDSE of Eq.~\eqref{theory:B:FullTDSE}, projecting onto the electronic states and neglecting the terms containing derivatives of the electronic states $\psi_i(\mathbf{r};R)$ with respect to the nuclear distance $R$, we obtain the following equation governing the nuclear motion
\begin{equation}
	i\partial_t \chi_i(t,R) = \sum_{j=1}^2 \Big( H^{(0)}_i \delta_{ij} +  V_{ij} \Big) \chi_j(t,R).
	\label{theory:B:NuclearTDSE}
\end{equation}
The Hamiltonian $H_i^{(0)} = T_N + E_i(R)$ describes the nuclear wave packet moving on the potential energy surface $E_i(R)$ and $V_{ij} = -F(t) D_{ij}(R)$ is the interaction with the field with $D_{ij}(R) = \braket{\psi_i(R) | \hat{D} | \psi_j (R)}$ the matrix element of the dipole operator integrated over the electronic coordinate. It should be noted that the two-lowest energy curves of the LiF molecule exhibit an avoided crossing at roughly $R=13.3$, where the neglecting of terms containing the derivatives $\partial_R \psi_i(\mathbf{r};R)$ is not valid. Since the initial nuclear wave packet is centered at $R\sim3$, and the ATA signal requires overlap between the excited- and ground state nuclear wave packet, the effect of the avoided crossing is negligible in the present study. 

After obtaining the nuclear wave functions, the expectation value of the dipole moment can be calculated as
\begin{equation}
	\braket{\hat{D}}_{\text{tot}}(t) = \sum_{i,j=1}^2 \int \! dR \, \chi_i^* (R,t) D_{ij}(R) \chi_{j}(R,t). \label{theory:B:dipole}
\end{equation}

In experimental settings, finite detector resolution and collisional broadening are effects that can lead to dephasing of the time-dependent dipole moment. A common approach \cite{wu2016theory,baekhoej2015,rorstad2017} is to impose a window function $W(t-\tau_0)$ in order to mimic this behavior in simulations. The function, which brings $\braket{\hat{D}}_{\text{tot}}(t)$ to zero over a given time interval, is defined as
\begin{equation}
W(t-\tau_0)=
\begin{dcases}
1 & (t<\tau_0)\\
\exp{\left[-\frac{(t-\tau_0)^2}{T_0^2/4} \right]}	& (\tau_0 \leq t),
\end{dcases} \label{theory:B:window}
\end{equation}
where the FWHM of the corresponding Gaussian is related to $T_0$ by $T_\text{FWHM} = \sqrt{\log{(2)}}T_0$, and $\tau_0$ is the starting time of the dephasing, which here is set to the end of the last pulse. The period is chosen to be long enough that the qualitative features we are interested in are not altered as the window function is imposed; in the present situation $T_0=100$ fs is an appropriate choice.

\subsection{\label{subsec:adiab}Fixed-nuclei adiabatic model}
In this section we derive an alternative method of obtaining ATAS spectra, meant to complement the numerical model of Sec.~\ref{subsec:TDSE}, as it allows for a deeper analytic investigation than its numerical counterpart. This method enables, in Sec.~\ref{sec:results}, the determination of the physical causes for individual features observed in the spectra by making a number of approximations, after which we arrive at relatively simple expressions for their response functions [Eq.~\eqref{theory:A:response_1}].

To capture the main features it is sufficient to consider the case of fixed nuclei, enabling the neglect of the term corresponding to nuclear kinetic energy, $T_N$, in Eq.~\eqref{theory:B:FullHamiltonian}. The internuclear distance $R$ at which the nuclei are frozen corresponds to the center of the ground state nuclear wave function at $R_0=3$; all quantities with $R$-dependence are thus evaluated at $R_0$. In the following we use the complex coefficients $a_i(t)$ in lieu of the nuclear wave functions $\chi_i(t,R)$ used in Sec.~\ref{subsec:TDSE}.

A basis of adiabatic states serves as a convenient foundation for analytical calculation, as seen in our previous investigations in He \cite{rorstad2017}. We express the state of the system as
\begin{equation}
\begin{aligned}
\ket{\Phi(t)}=&\,a_1(t)\ket{\phi_1(t)}e^{-i\int_{t_0}^t\!dt'E_{1a}(t')}\\
& + a_2(t)\ket{\phi_2(t)}e^{-i\int_{t_0}^t\!dt'E_{2a}(t')}, \label{theory:C:system}
\end{aligned}
\end{equation}
where $\ket{\phi_i(t)}$ are adiabatic states with time-dependent energies $E_{ia}(t)$. The time $t_0$ in Eq.~\eqref{theory:C:system} is chosen so that it precedes the start of the NIR pulse. The adiabatic states are defined by
\begin{equation}
[ T_e+V - F_\text{NIR}(t)\hat{D} ]\ket{\phi_i(t)} = E_{ia}(t)\ket{\phi_i(t)}\label{theory:C:adiabcond},
\end{equation}
where $V=V_{NN}+V_{ee}+V_{Ne}$.

We seek to obtain $\braket{\Phi(t)|\hat{D}|\Phi(t)}$ (now calculated without integration over $R$), which requires the determination of the unknown quantities of Eq.~\eqref{theory:C:system}. In light of Eq.~\eqref{theory:C:adiabcond} the adiabatic states and corresponding energies can be found, in a basis of the field free states ($\ket{\psi_1},\ket{\psi_2}$), as the eigenstates and eigenenergies of the following matrix at a given instant of time $t$
\begin{equation}
\begin{bmatrix}
E_1(R_0) - F_\text{NIR}(t)D_{11}(R_0) 	& -F_\text{NIR}(t)D_{12}(R_0) \\
-F_\text{NIR}(t)D_{12}(R_0) 			& E_2(R_0) - F_\text{NIR}(t)D_{22}(R_0) \label{theory:C:mat}
\end{bmatrix},
\end{equation}
where we have used that $(T_e+V)\ket{\psi_i}=E_i(R_0)\ket{\psi_i}$. The  normalized eigenstates can be expressed as $\ket{\phi_i(t)}=\alpha_{i1}(t)\ket{\psi_1} + \alpha_{i2}(t)\ket{\psi_2}$, in which case we have $\alpha_{i1}^2(t) + \alpha_{i2}^2(t) = 1$ per definition. In the following we suppress dependencies in the notation for brevity where appropriate, and we use dotted variables to indicate differentiation with respect to $t$.

In order to determine the time-dependent values of the complex coefficients $a_i(t)$, we insert Eq.~\eqref{theory:C:system} into the TDSE, and project the resulting equation onto each of the adiabatic states $\bra{\phi_i}$, obtaining two coupled equations 
\begin{align}
\dot{a}_1=&\,ia_1F_\text{UV}\braket{\phi_1|\hat{D}|\phi_1}\nonumber\\
\,&-a_2\braket{\phi_1|\dot{\phi}_2}e^{-i\int_{t_0}^t\!dt'(E_{2a}-E_{1a})}\label{theory:C:adot1}\\
\,&+ia_2F_\text{UV}\braket{\phi_1|\hat{D}|\phi_2} e^{-i\int_{t_0}^t\!dt'(E_{2a}-E_{1a})} \nonumber\\ 
\dot{a}_2=&\,iF_\text{UV}a_2\braket{\phi_2|\hat{D}|\phi_2} \nonumber\\
\,&-a_1\braket{\phi_2|\dot{\phi}_1} e^{-i\int_{t_0}^t\!dt'(E_{1a}-E_{2a})} \label{theory:C:adot2}\\
&+iF_\text{UV}a_1\braket{\phi_2|\hat{D}|\phi_1} e^{-i\int_{t_0}^t\!dt'(E_{1a}-E_{2a})} , \nonumber
\end{align}
where terms containing $\braket{\phi_i|\dot{\phi}_i}$ have been dropped, since
\begin{equation}
\begin{aligned}
\braket{\phi_i|\dot{\phi}_i} &= \alpha_{i1}\dot{\alpha}_{i1} + \alpha_{i2}\dot{\alpha}_{i2}\\
&=\frac{1}{2}\frac{d}{dt}(\alpha_{i1}^2 + \alpha_{i2}^2), \label{theory:C:normaldrop}
\end{aligned}
\end{equation}
which evidently is zero given the normalization of the eigenstates. The amplitudes $a_1(t)$ and $a_2(t)$ are obtained by solving Eqs.~\eqref{theory:C:adot1}-\eqref{theory:C:adot2} numerically, with initial values $a_1(t_0)=1$ and $a_2(t_0)=0$. 

In the adiabatic model, the expectation value for the time-dependent dipole moment is
\begin{align}
\braket{\Phi|\hat{D}|\Phi} \approx&\, |a_1|^2\braket{\phi_1|\hat{D}|\phi_1}\label{theory:C:dipmom}\\
&+2\,\text{Re}\, \left[a_1^*a_2\braket{\phi_1|\hat{D}|\phi_2}e^{-i\int_{t_0}^t\!dt'(E_{2a}-E_{1a})}\right]\nonumber,
\end{align}
where, given modestly intense fields, and due to the fact that we start out the in ground state, the factor $|a_2|^2$ is relatively small and the term in which it appears is neglected. The model can then be realized by inserting the determined quantities into Eq.~\eqref{theory:C:dipmom}, Fourier transforming the dipole moment, and inserting it into Eq.~\eqref{theory:A:response_1} along with the Fourier transformed UV pulse. 

The features which we are particularly focused on in this paper are the ladder and the LISs (see Sec.~\ref{sec:results}), and they differ significantly in both character and origin. How they arise from the adiabatic model derived in this section, the interpretation of the underlying physical processes, and the properties of these features are investigated further in Sec.~\ref{subsec:features}, where we consider the LISs and the ladder structure separately---with the adiabatic model used as the starting point for an analytical investigation.

\section{\label{sec:results}Results}

In this section, we exemplify the theory of Sec.~\ref{section:theory} with the polar diatomic LiF molecule. After considering spectra obtained through fully numerical calculations, we follow up with analytical and semi-analytical investigations based on the adiabatic method---elucidating origins and properties of characteristic features in the spectra. Finally, we examine the effects of having target molecules that are either oriented or aligned with respect to the incoming electric field. 

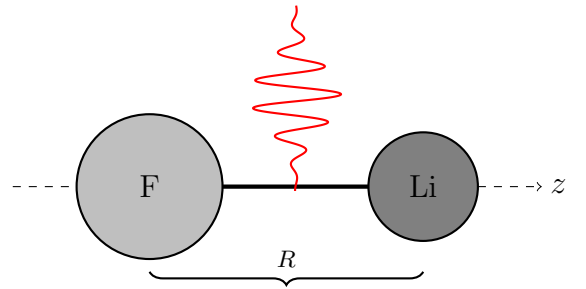
\begin{figure}
 \begin{tikzpicture}[scale=1.2]
 	\draw [fill=lightgray, thick] (0,0)  circle [radius=0.8];
 	\node at (0,0) {\large F};
 	\draw [fill=gray, thick] (3,0) circle [radius=0.6];
 	\node at (3,0) {\large Li};
 	\draw [ultra thick] (0.8,0) -- (2.4,0);
 	\draw [red, thick, domain=2:-0.05] plot[samples=500] ({0.5*sin(20*\x r)*exp(-4*pow((\x-1),2)) + 1.6}, {\x});
 	\draw [dashed] (-1.5,0) -- (-0.8,0);
 	\draw [dashed] [->] (3.6,0) -- (4.3,0);
 	\node [right] at (4.3,0) {\large $z$};
 	\draw [decorate,decoration={brace,amplitude=5pt,mirror},xshift=0,yshift=-4pt,thick] (0,-0.8) -- (3,-0.8)
 		node [midway,yshift=5pt] {$R$};
 \end{tikzpicture}
\caption{\label{fig:molecule} Illustration of LiF molecule and incoming NIR field. The molecule is oriented in the $z$-direction, which is parallel to the polarization of the UV and the NIR field. $R$ is the internuclear distance.}
\end{figure}

\begin{figure}
\includegraphics[width=0.48\textwidth]{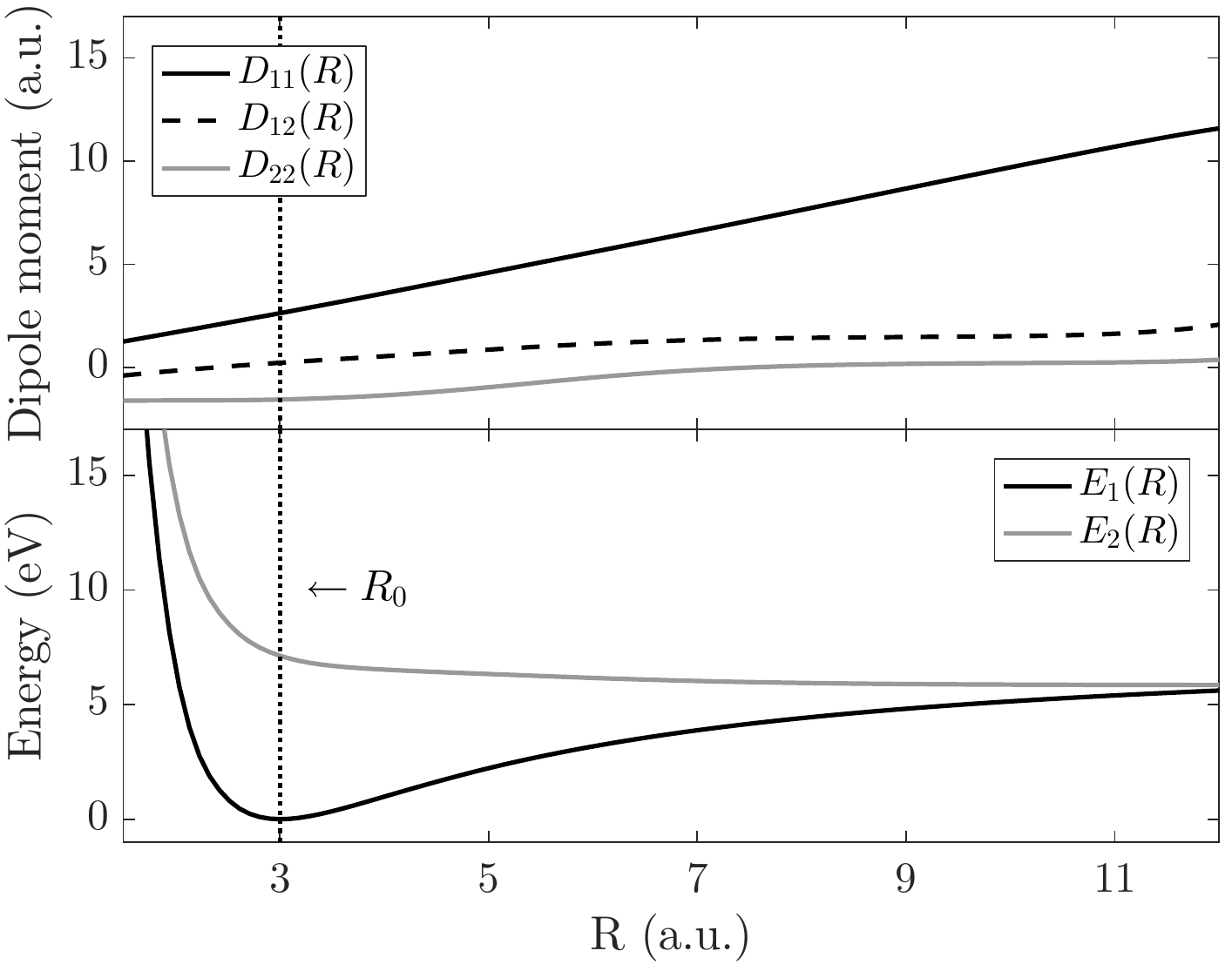}
\caption{\label{fig:system} Dipole moments (top panel) and potential energy surfaces (bottom panel) of LiF, as functions of the internuclear distance $R$. Dotted line indicates center of ground state nuclear wave function at $R_0=3$, where $E_1(R_0)=0$~eV, $E_2(R_0)=7.07$~eV, $D_{11}(R_0)=2.62$, $D_{12}(R_0)=0.22$, and $D_{22}(R_0)=-1.55$. Curves are interpolated from data in Ref.~\cite{werner1981}.}
\end{figure}

An illustration of the LiF molecule is shown in Fig.~\ref{fig:molecule}, along with the incoming field, with the molecule oriented along the $z$-axis parallel to the field polarization. As mentioned in Sec.~\ref{subsec:TDSE} we expand the wave function in the two lowest Born-Oppenheimer states and propagate the nuclear wave functions on the potential energy curves. The two lowest potential energy curves, $E_{1}(R)$ and $E_{2}(R)$ of LiF are shown in Fig.~\ref{fig:system} together with the associated dipole moments, $D_{11}(R)$, $D_{12}(R)$ and $D_{22}(R)$; the dipole moments used were interpolated from data given in Ref.~\cite{werner1981}. An important characteristic of the excited curve $E_{2}(R)$ is its slope at $R=3$, which ensures that any wave packet excited by the UV pulse to this curve will quickly propagate towards larger $R$. This effectively cuts off the dipole interaction between the ground state and excited state wave packets  at a time scale shorter than typical dephasing time scales. This crucial effect is automatically included in the full numerical calculations of Eq.~\eqref{theory:B:NuclearTDSE}, but under an assumption of fixed nuclei, as used in the section above (Sec.~\ref{subsec:adiab}), it is not. Imposing on the excited state population a window function $W_N(t-\tau)$ with $T_{0,N}=3.6$ fs related to the time it takes for the excited wave packet to depart from the range of $R$ in which it overlaps with the ground state wave packet, ensures that the effect is properly imitated.

The critical distinction of the current system compared with systems previously explored with ATAS, is the presence of a nonzero permanent dipole. The concept of dipole transitions that are disallowed due to identical parity between initial and final states does not apply to systems in which states do not posses a definite parity; the notion of a dark state \cite{baekhoej2015_2} with respect to transitions from the ground state (i.e. single-photon absorption between the states prohibited) is in this context unsuitable. In ATAS this has important consequences, and we consider them in detail.

\subsection{\label{subsec:results:TDSE}Full numerical model}

The nuclear wave packets of the full model, Eq.~\eqref{theory:B:NuclearTDSE}, are propagated using a split step method with time step $\Delta t = 0.05$ (a.u.), box size $R_{\text{max}} = 25$ (a.u.), and grid size $\Delta R = R_{\text{max}}/N_R$ with $N_R = 1024$ and with a complex absorbing potential at the boundary to remove unphysical reflections. The initial ground state is found by imaginary time propagation without the fields. 

\begin{figure*}
\includegraphics[width=0.9\textwidth]{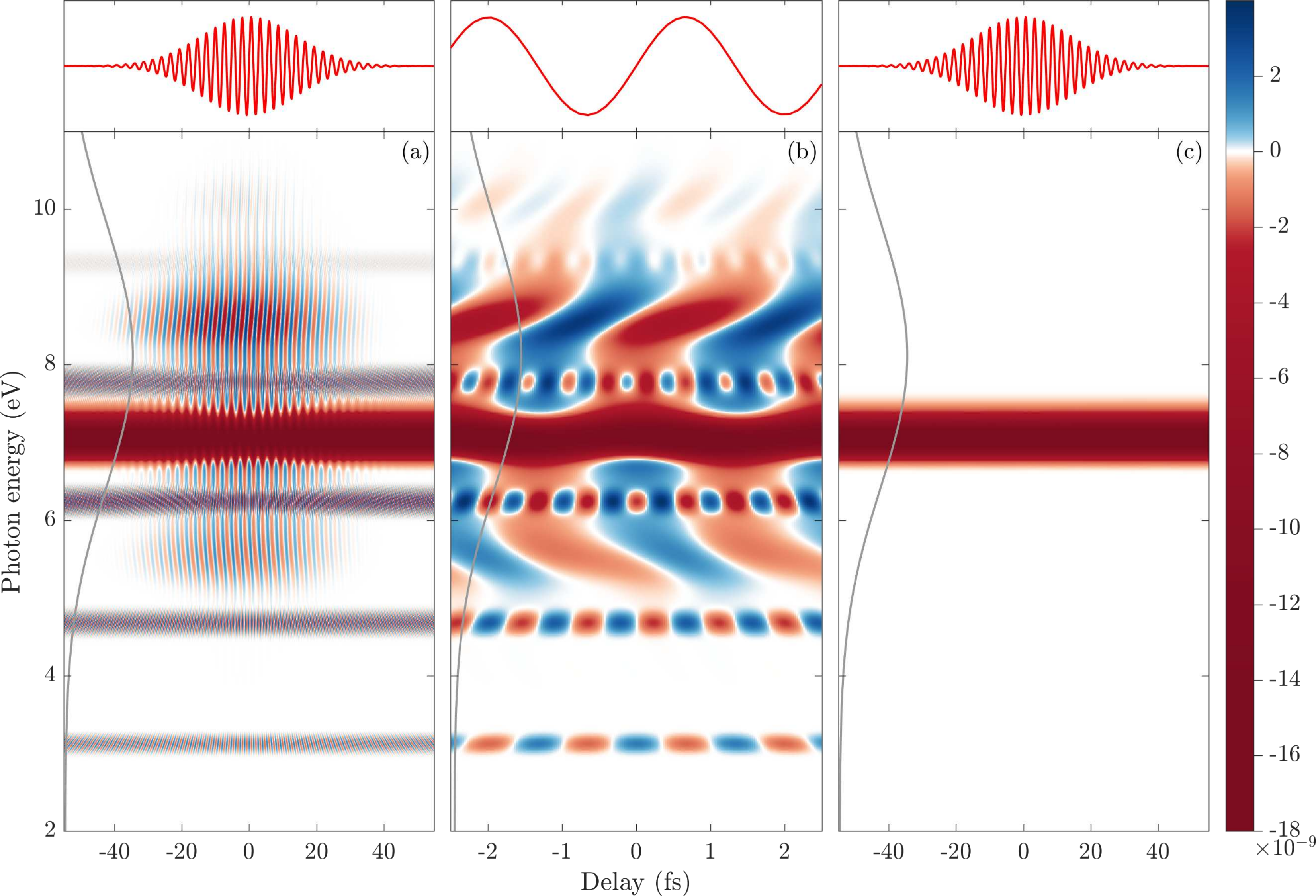}
\caption{\label{fig:intro} Attosecond transient absorption spectrum of the LiF molecule as calculated by Eq.~\eqref{theory:A:response_1}, exhibiting the difference between systems with a permanent dipole moment (e.g. polar molecules) in panels~(a) and (b), and systems with no permanent dipole moment (e.g. non-polar molecules) in panel~(c). The main features visible in panel (a) include the main absorption line at $E_2(R_0)=7.07$~eV; light-induced structures centered at energies $E_2(R_0)\pm\omega_\text{NIR}$ and $E_2(R_0)+2\omega_\text{NIR}$, i.e. at $5.51$~eV, $8.61$~eV, and $10.16$~eV; and the rungs of the ladder structure at $E_1(R_0)=n\omega_\text{NIR}$, i.e. at $n$ times $1.55$~eV, with $n=\{2,3,4,5,6\}$. Panel~(b) exhibits a zoomed in version of panel~(a), which yields a clearer view of the characteristics of the ladder structure. In panel~(c) the main absorption line is the only feature remaining. The frequency bandwidth of the UV pulse is shown in gray in each figure. The top panels depict the NIR pulse centered at $\tau=0$~fs. The color scale on the right displays the signal strength in arbitrary units. The pulse parameters are given in the text following Eq.~\eqref{theory:A:pulse}.}
\end{figure*}

Figure~\ref{fig:intro} shows the ATA spectrum of the LiF molecule calculated by Eq.~\eqref{theory:A:response_1} using the full numerical model.
In Fig.~\ref{fig:intro}~(a) we can see the broad main absorption line centered at $7.07$~eV, corresponding to a vertical Franck-Condon transition at $R = R_0$ from the ground state curve $E_1$ to the excited state curve $E_2$. Which-way interference due to multiple pathways of population transfer sharing a final destination \cite{chen2013} can be observed at the top and bottom of the main absorption line for delays between roughly $-40<\tau<40$. The pathways correspond to absorption of a single UV photon, or absorption of one UV photon and then either absorption of one NIR photon or emission of one or two NIR photons.  The LISs are also signatures of these pathways, and can be seen centered around $E_2(R_0)\pm\omega_\text{NIR}$ and $E_2(R_0)+2\omega_\text{NIR}$. Finally, due to the presence of the permanent dipole, we observe a new ladder structure at energies $E_1(R_0)=n\omega_\text{NIR}$. Both the LISs and the ladder structure will be discussed further in the following section where we derive analytical and semi-analytical expressions. Figure~\ref{fig:intro}~(b) shows a zoomed in view of the spectrum in Fig.~\ref{fig:intro}~(a), giving a clearer picture of the structure of the ladder. Figure~\ref{fig:intro}~(c) shows the spectrum that would arise if the molecule had no permanent dipole, i.e. if $D_{11} = D_{22} = 0 $. 

Apart from the LISs and the ladder, which comprise the main focus of this paper, some of the typical features that were described briefly in the introduction are notably absent from Fig.~\ref{fig:intro}. As mention previously, the nuclear wave packet belonging to the electronic excited state interacts only briefly with the ground state wave packet before propagating towards greater $R$. Thus, any features that require the presence of an enduring dipole interaction between the ground- and excited state populations will not be generated, explaining the absence of a certain species of oscillating fringes \cite{chen2013} and sidebands \cite{rorstad2017} along the main absorption line. 

\subsection{\label{subsec:features}Features} 

\subsubsection{\label{subsubsec:LIS}Light-induced structures}
In systems with no permanent dipole, the presence of LISs is an indicator of a two-photon process in which one UV photon has been absorbed and one NIR photon has either been absorbed or emitted, depending on the location of the feature, in a two-photon transition from the ground state to a dark state \cite{chen2012, baekhoej2015_2}. With a nonexistent permanent dipole there is no visible signature in the spectrum at the energy of the dark state itself, since in the absence of a dipole coupling between the dark state and ground state an oscillating dipole moment cannot be established. 
 
\begin{figure}
\includegraphics[width=0.47\textwidth]{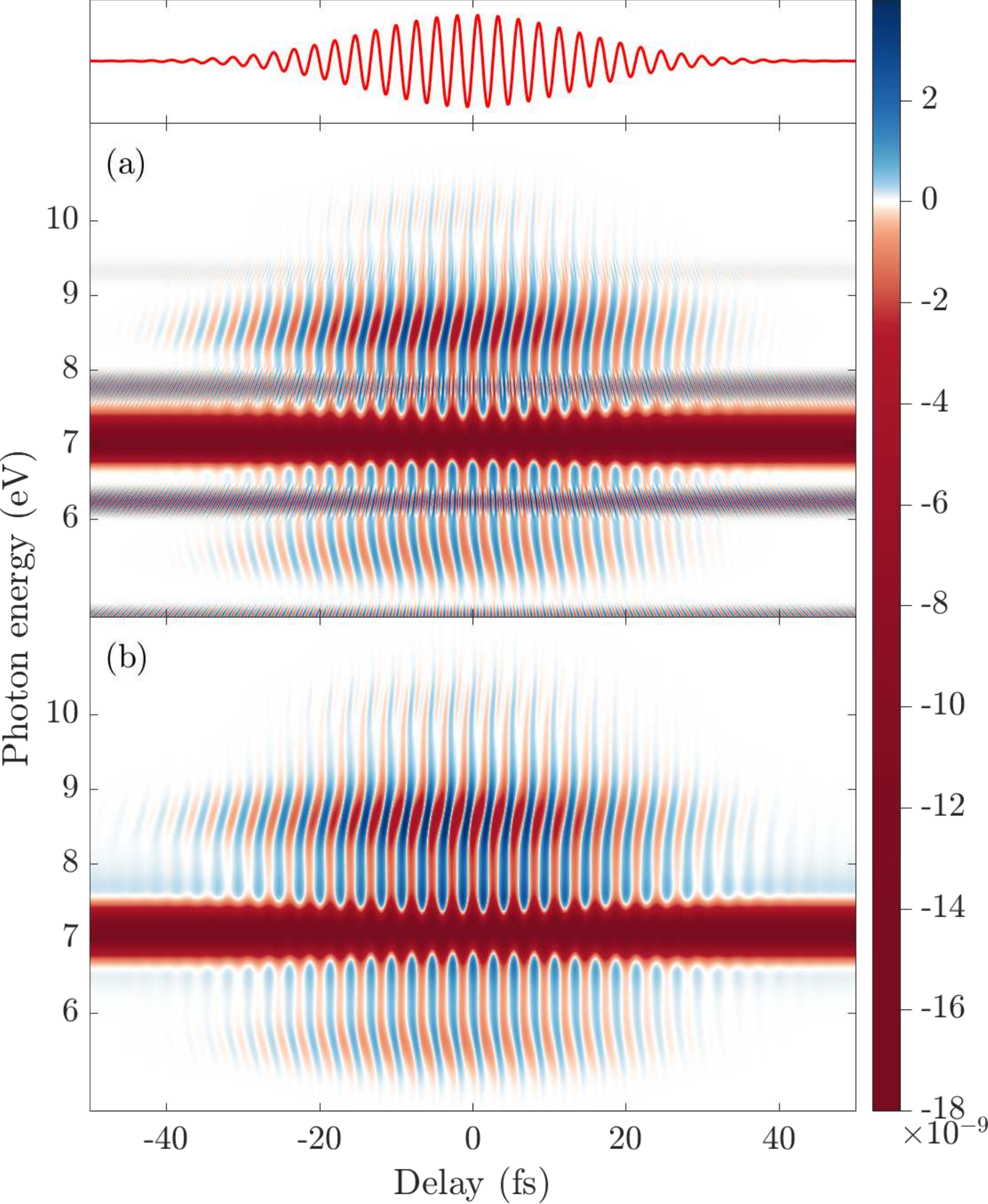}
\caption{\label{fig:LIS} Attosecond transient absorption spectrum of the LiF molecule as calculated by Eq.~\eqref{theory:A:response_1}, exhibiting the LISs. In panel~(a) the LISs from the full numerical method of Sec.~\ref{subsec:TDSE} are shown. Panel~(b) displays the LISs corresponding to the time-dependent dipole moment of Eq.~\eqref{features:A:dip3}, with phases from Eq.~\eqref{features:A:phase}. The top panel depicts the NIR pulse centered at $\tau=0$~fs. The color scale on the right displays the signal strength in arbitrary units. The pulse parameters are given in the text following Eq.~\eqref{theory:A:pulse}.}
\end{figure} 

Figure~\ref{fig:LIS}~(a) displays the ATAS spectrum obtained through the full numerical method described in Sec.~\ref{subsec:TDSE}, with a focus on the LISs. Unlike LISs observed in the past, they surround a visible absorption line, corresponding to population transfer from the ground state to a bright state at $E_2(R_0)=7.07$~eV. Two LISs can be seen centered at energies $E_2(R_0)-\omega_\text{NIR}=5.51$~eV and $E_2(R_0)+\omega_\text{NIR}=8.61$~eV, and a third, weaker one, at $E_2(R_0)+2\omega_\text{NIR}=10.16$~eV; the latter can be interpreted as a sign of the absorption of one UV photon and the emission of two NIR photons. 

To improve the description of the LISs, a series of approximations can be applied to the adiabatic model of Sec.~\ref{subsec:adiab}, leaving only the parts most important in the generation of the LISs. The first step in this process is to isolate the part of the time-dependent dipole moment in Eq.~\eqref{theory:C:dipmom} which is responsible for the LISs. The phase factor $\exp \! {\left[-i\int_{t_0}^t\!dt'(E_{2a}-E_{1a})\right]}$ in the second term implies a shift to energies around $E_2(R_0)$ when taking Fourier transform, suggesting relevance for the LISs. By neglecting the first term of Eq.~\eqref{theory:C:dipmom}, we are left with
\begin{equation}
\braket{\hat{D}}_\text{LIS}=2\,\text{Re}\,\left[a_1^*a_2\braket{\phi_1|\hat{D}|\phi_2}e^{-i\int_{t_0}^t\!dt'(E_{2a}-E_{1a})}\right]. \label{features:A:dip1}
\end{equation}

We start by considering the coefficients $a_1(t)$, $a_2(t)$. First we assume that $a_1(t)$ changes little throughout the interaction with the field, relative to its initial value, which amounts to setting $a_1(t)=1$. The creation of the LISs observed in our spectra involves the NIR field to either first or second order; so we posit that up to second order in the field the change in the adiabatic state $\ket{\phi_1}$ is sufficiently slow to justify the neglecting of the non-adiabatic term $\braket{\phi_2|\dot{\phi}_1}$ in Eq.~\eqref{theory:C:adot2}. Both remaining terms in Eq.~\eqref{theory:C:adot2} contain the factor $F_\text{UV}$, which is nonzero only at a brief interval centered at the time $t=\tau$. Given the short integration time, we assume that time-dependent parts of the energies and the coupling factor in Eq.~\eqref{theory:C:adot2} are of less importance relative to the constant parts, and can be neglected, yielding
\begin{align}
\braket{\phi_2|\hat{D}|\phi_1}&\approx D_{12}\label{features:A:approx1}\\ 
e^{-i\int_{t_0}^t\!dt'(E_{1a}-E_{2a})}&\approx e^{-i(t-t_0)(E_1-E_2)},
\end{align}
which amounts to a reduction to the equivalent field free quantities. Another significant assumption is that the exponential factor in Eq.~\eqref{features:A:dip1} is the main contributing factor to the LISs, and so we also impose the approximation from Eq.~\eqref{features:A:approx1} on Eq.~\eqref{features:A:dip1}. 

Given the weak intensity of the UV field, we can treat its interaction with the system perturbatively. To zeroth order in the UV field the coefficients remain unchanged, and we have $a_2^{(0)}(t)=0$, where the superscript denotes the order of approximation. We obtain the first order approximation, $a_2^{(1)}(t)$, by inserting $a_1(t)=1$ and $a_2^{(0)}(t)=0$ into Eq.~\eqref{theory:C:adot2} and integrating
\begin{equation}
a_2^{(1)} = iD_{12}W_N(t-\tau)\int_{t_0}^t\!dt'\,F_\text{UV}e^{-i(t'-t_0)(E_1-E_2)}, \label{features:A:dip2}
\end{equation}
where $W_N(t-\tau)$ is the window function described in the text at the start of Sec.~\ref{sec:results}, which reproduces the abrupt suppression of the dipole moment due to dissociating wave packets in models with free nuclei. 

Returning to Eq.~\eqref{features:A:dip1}, we now have
\begin{equation}
\braket{\hat{D}}_\text{LIS}=2D_{12}\,\text{Re}\,\left[a_2^{(1)}e^{-i\int_{t_0}^t\!dt'(E_{2a}-E_{1a})}\right], \label{features:A:dip3}
\end{equation}
and we turn our attention to the phase factor. Taylor expanding the instantaneous eigenvalues of matrix~\eqref{theory:C:mat} about $F_\text{NIR}=0$ and keeping only terms up to second order, we have
\begin{align}
E_{1a}&=E_1 - D_{11}F_\text{NIR}\label{features:A:E1} - \frac{D_{12}^2}{E_2 - E_1}F^2_{\text{NIR}}   \\ 
E_{2a}&=E_2 - D_{22}F_\text{NIR}\label{features:A:E2} + \frac{D_{12}^2}{E_2 - E_1}F^2_{\text{NIR}}.
\end{align}
In light of Eqs.~\eqref{features:A:E1} and \eqref{features:A:E2}, by factoring out the constant part of the integrand in the phase factor, expressing the rest in terms of its Taylor series, and neglecting terms of third order or higher in the field $F_\text{NIR}$, we obtain
\begin{equation}
\begin{aligned}
e^{-i\int_{t_0}^t\!dt'(E_{2a}-E_{1a})} \approx &\,e^{-i(t-t_0)(E_2-E_1)}e^{i\int_{t_0}^t\!dt'(D_{22}-D_{11})F_\text{NIR}}\\
\approx&\,  \,e^{-i(t-t_0)(E_2-E_1)}\\
&\,\times\bigg[1+i(D_{22}-D_{11})\int_{t_0}^t\!dt'F_\text{NIR}\\
&\,- \frac{1}{2}(D_{22}-D_{11})^2\left(\int_{t_0}^t\!dt'F_\text{NIR}\right)^{\!\!2} \bigg], \label{features:A:phase}
\end{aligned} 
\end{equation}
where we have neglected an additional second order term due to it being more than an order of magnitude smaller.

The simplified model of the LISs is finally obtained by inserting Eq~\eqref{features:A:phase} into Eq.~\eqref{features:A:dip3}, taking the Fourier transform of $\tilde{D}_\text{LIS}$, and inserting into Eq.~\eqref{theory:A:response_1}. The resulting spectrogram is shown in Fig.~\ref{fig:LIS}~(b). The absorption line and three LISs match the corresponding features found by the full TDSE calculation in Fig.~\ref{fig:LIS}~(a). 

Equations~\eqref{features:A:dip2}, \eqref{features:A:dip3}, and \eqref{features:A:phase} give rise to the LISs, and from them we can surmise the origins and properties of the LISs. Evidently, both the off-diagonal ($D_{12})$ and diagonal ($D_{11}$, $D_{22}$) dipole matrix elements are essential. $D_{12}\neq0$ is a necessary condition for the presence of any features at all, with the overall signal strength proportional to $D_{12}^2$, as seen by combining Eqs.~\eqref{features:A:dip3} and \eqref{features:A:phase}. The difference between the permanent dipole elements $D_{22}-D_{11}$ influences the signal strength of the LISs, but if they are set to zero the main absorption line will remain, in agreement with the findings demonstrated in Fig.~\ref{fig:intro}~(c). The presence and qualitative features of the LISs originate in the field-induced time-dependent adiabatic energies. In our previous work on ATAS in He \cite{rorstad2017}, where a similar adiabatic model was employed, the LISs were found to originate in the term corresponding to mixing of states due to an infrared field, comparable to the variable $\alpha_{12}(t)$ defined after the matrix~\eqref{theory:C:mat}. This suggests that the LISs cannot consistently be attributed to one factor across different systems, without taking into account the specific properties of the system. 

\subsubsection{\label{subsubsec:ladder}Ladder structures}
\begin{figure*}
\includegraphics[width=0.7\textwidth]{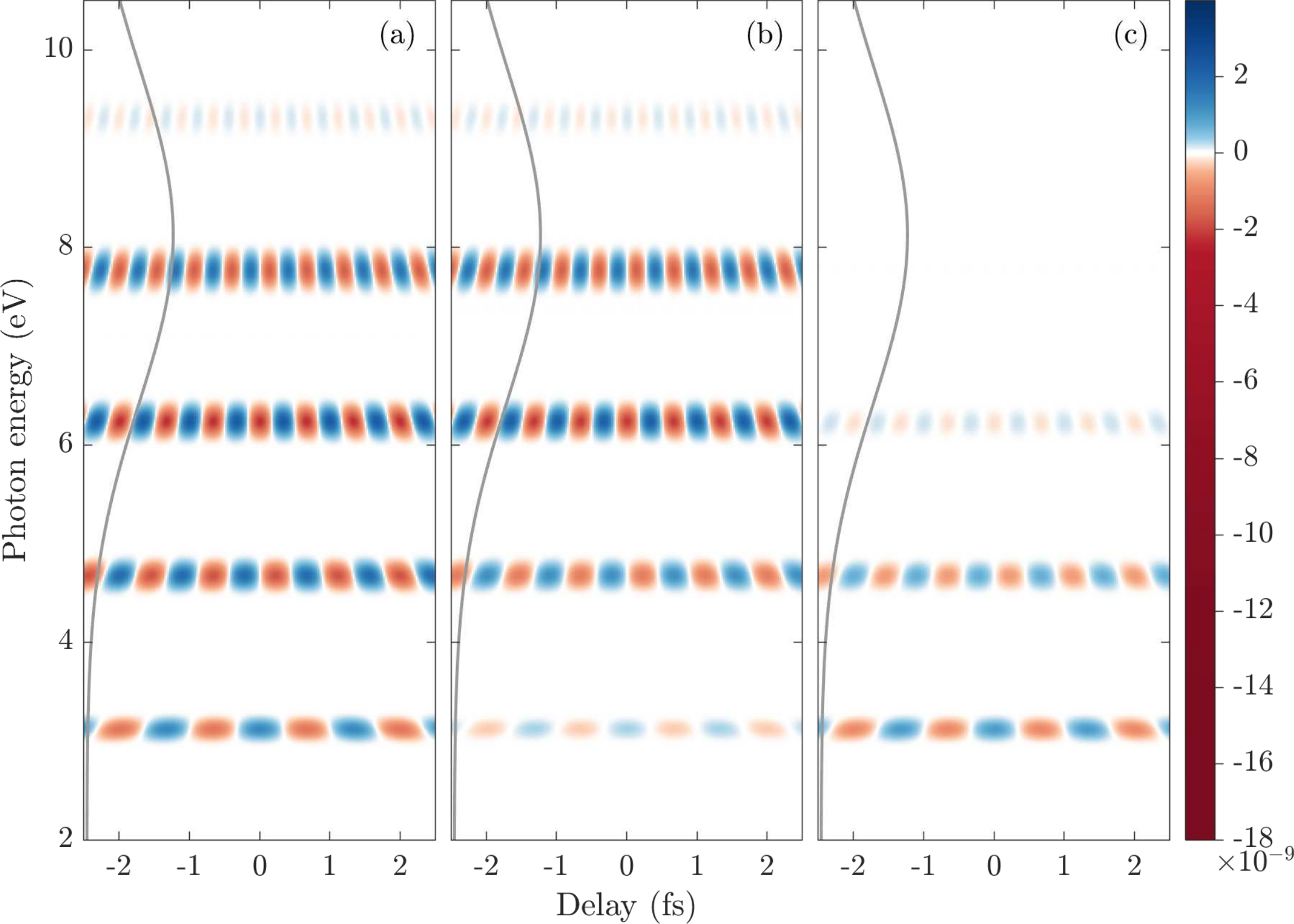}
\caption{\label{fig:ladder} Attosecond transient absorption spectrum of the LiF molecule as calculated by Eq.~\eqref{theory:A:response_1}, exhibiting an isolated view of the ladder feature. In panel~(a) the ladder as calculated by the full adiabatic method of Sec.~\ref{subsec:adiab} is shown. In panel~(b) the non-adiabatic component of the ladder, corresponding to the second term of Eq.~\eqref{features:B:dip1} is displayed. In panel~(c) the adiabatic component of the ladder is shown, corresponding to the first term of Eq.~\eqref{features:B:dip1}, which is indistinguishable from the spectrum derived from the analytical expression of Eq.~\eqref{features:B:analytical}. The frequency bandwidth of the UV pulse is shown in gray in each figure. The color scale on the right displays the signal strength in arbitrary units. The pulse parameters are given in the text following Eq.~\eqref{theory:A:pulse}.}
\end{figure*}

In the previous section we derived a simplified model for the LISs. Here we will do the same for the ladder structures, culminating in an analytic expression for the response function, which reproduces certain parts of the ladder feature. The isolated ladder feature as calculated by the full adiabatic model derived in Sec.~\ref{subsec:adiab} is shown in Fig.~\ref{fig:ladder}~(a). It is nearly indistinguishable from the ladder as calculated by the full numerical method of Sec.~\ref{subsec:TDSE}, shown in Fig.~\ref{fig:intro}~(c), the only distinction being a very slight difference in overall signal strength. The ladder feature has not been observed previously in ATA spectrograms, and arises as a result of a nonzero permanent dipole. The rungs of the ladder are located at energies $E_1(R_0)+n\,\omega_\text{NIR}$ ($n=\{1,2,3,4,5,6\}$), and the $n$'th rung oscillates with the delay $\tau$ as $n\,\omega_\text{NIR}$. Note that the first rung is not included in our figures, because such low frequency components are not practically obtainable in the standard ATAS experimental setup. The full frequency bandwith of the NIR pulse is filtered out after interaction with the target, meaning that only the modulated UV field is incident upon the detector (see, for example, Ref.~\cite{beck2015probing}). For UV pulses with a relatively low central frequency, however, a possible consequence is that the lower tail end of the frequency bandwidth could also be filtered out (see gray lines in Figs.~\ref{fig:intro}, \ref{fig:ladder} or \ref{fig:ovsa} for the UV pulse in the frequency domain). 

The derivation of the simplified expression for the ladder starts by noting that in ATAS, the UV field $F_\text{UV}$ plays two roles. First, it is a factor in Eq.~\eqref{theory:A:response_1}, reflecting the fact that the response function is an expression of the interference between the incoming UV field and the dipole response of the target. Second, it enters into the calculations of the dipole moment of the system itself; in the adiabatic model of Sec.~\ref{subsec:adiab} it is present in Eqs.~\eqref{theory:C:adot1} and \eqref{theory:C:adot2}. The latter role of the UV pulse is not involved in the generation of the ladder feature, and we can therefore neglect $F_\text{UV}$ when calculating $a_1(t)$ and $a_2(t)$. This is precisely the way in which we isolate the ladder in Fig.~\ref{fig:ladder}, as the other features require a nonzero $F_\text{UV}$ in Eqs.~\eqref{theory:C:adot1} and \eqref{theory:C:adot2}. As in Sec.~\ref{subsubsec:LIS} [see text following Eq.~\eqref{features:A:dip1}], we assume that $a_1(t)=1$, so that Eq.~\eqref{theory:C:adot2} now reads
\begin{equation}
\dot{a_2}=-\braket{\phi_2|\dot{\phi}_1}e^{-i\int_{t_0}^t\!dt'(E_{1a}-E_{2a})} \label{features:B:adot2}.
\end{equation}
and the dipole moment is
\begin{equation}
\begin{aligned}
\braket{\hat{D}}_\text{ladder} =&\, \braket{\phi_1|\hat{D}|\phi_1} \\
&\,+ 2\,\text{Re}\,\left[a_2\braket{\phi_1|\hat{D}|\phi_2}e^{-i\int_{t_0}^t\!dt'(E_{2a}-E_{1a})} \right].\label{features:B:dip1}
\end{aligned}
\end{equation}

The two terms in Eq.~\eqref{features:B:dip1} each contribute to the generation of the full ladder feature. In Sec.~\ref{subsubsec:LIS} we argued for neglecting of the small terms involving $\braket{\phi_1|\dot{\phi}_2}$ and $\braket{\phi_2|\dot{\phi}_1}$ since we knew that the process behind the LISs involved the NIR field only up to second order. The appearance of the ladder feature, with rungs extending up to $E=6\,\omega_\text{NIR}$, suggests the involvement of the NIR field to at least sixth order, and the same approximation does not hold. In fact, the contribution from the second term to the ladder, seen in Fig.~\ref{fig:ladder}~(b) is greater than the contribution from the first term, seen in Fig.~\ref{fig:ladder}~(c). This is a clear instance of the adiabatic condition breaking down, where the adiabatic states $\ket{\phi_i}$ are no longer varying slowly over time relative to the dynamics of the system. 

Simplification or exclusion of any factors involved in Eq.~\eqref{features:B:adot2} or in the second term of Eq.~\eqref{features:B:dip1} entails significant alterations to the ladder, implying that the full ladder structure is not attributable to a single factor, but rather a complicated combination of factors, including the field-induced phases $\exp{\left[-i\int_{t_0}^t\!dt'(E_{2a}-E_{1a})\right]}$ and the dipole coupling between the adiabatic states $\braket{\phi_1|\hat{D}|\phi_2}$. The complicated character of these terms suggests that getting a clear understanding of the contribution from each of its factors is unrealistic. Instead we focus on the relatively uncomplicated first term, $\braket{\phi_1|\hat{D}|\phi_1}$, whose contribution to the spectrogram can be captured in a fully analytical expression.

The $\braket{\phi_1|\hat{D}|\phi_1}$ term represents the adiabatic contribution to the ladder structure, and more specifically it is involved in the generation of the rungs located at energies $E=n\,\omega_\text{NIR}$ where $n=\{2,3,4\}$, as seen in Fig.~\ref{fig:ladder}~(c). To proceed with the derivation, we Taylor expand $\braket{\phi_1|\hat{D}|\phi_1}$ in orders of the NIR field. The location of the rungs of the ladder suggests that we should include terms up to fourth order, which by Fourier transform translates to shifts in energy up to $4\times\omega_\text{NIR}$. In general, each coefficient of the Taylor expansion is a sum of several terms; by keeping only the dominating terms for each coefficient, the expansion can be shown to be  approximately
\begin{equation}
\braket{\phi_1|\hat{D}|\phi_1}\approx D_{12}^2\sum_{n=2}^4 (n+1)\frac{(D_{22}-D_{11})^{n-1}}{(E_2-E_1)^n}F_\text{NIR}^n. \label{features:B:dip2}
\end{equation} 

What remains to obtain a fully analytical expression for the response function [Eq.~\eqref{theory:A:response_1}] is then to obtain the Fourier transformed UV field, the derivation of which can be found in Appendix~\ref{section:appendix_UV}, and Fourier transforming Eq.~\eqref{features:B:dip2}, amounting to calculating the Fourier transform of $F_\text{NIR}^n$ up to $n=4$, which we relegate to Appendix~\ref{section:appendix_IR}. Combining the final expressions from the appendices, Eqs.~\eqref{app:A:final} and \eqref{app:B:final}, with Eqs.~\eqref{features:B:dip2} and inserting into Eq.~\eqref{theory:A:response_1} yields the response function for the adiabatic part of the ladder structure
\begin{equation}
\begin{aligned} \label{features:B:analytical}
\tilde{S}(\omega,\tau)=&\,\frac{\pi\rho D_{12}^2 T_\text{NIR} T_\text{UV}F_{0,\text{UV}}}{4c}\\
&\times\frac{\omega^2}{\omega_\text{UV}}\exp{\left[ -\frac{T_\text{UV}^2}{16}(\omega-\omega_\text{UV})^2\right]}\\
&\times\sum\limits_{n=2}^4(n+1)\frac{(D_{22}-D_{11})^{n-1}}{(E_2-E_1)^n}\frac{F_{0,\text{NIR}}^n}{\sqrt{n}}\\ 
&\times\exp{\left[-\frac{T_\text{NIR}^2}{16n}(\omega-n\omega_\text{NIR})^2\right]}\left(-\frac{1}{2}\right)^n\\
&\times\text{Im}\left\{i^{n+1}\exp{(i\tau\omega)}\exp{[i(n-1)\varphi]}\right\},
\end{aligned}
\end{equation}
which, with $\varphi=0$, exactly reproduces the spectrogram of Fig.~\ref{fig:ladder}~(c). 

The advantages of having a fully analytic expression include that the origin of every characteristic property of the feature can be deduced, and that the various dependencies of the feature are explicitly demonstrated. Equation~\eqref{features:B:analytical} represents the adiabatic contribution to the ladder structure, which for the current system is secondary to the non-adiabatic contribution. This could change, however, for example in systems interacting with a more slowly varying field, e.g. an infrared pulse at a lower frequency. The main contribution to the adiabatic part of the ladder feature comes from the $\braket{\phi_1|\hat{D}|\phi_1}$ term of Eq.~\eqref{features:B:dip1}, which represents a mixing of the field free states $\ket{\psi_1}$ and $\ket{\psi_2}$. The $n$-dependent, or rung-dependent, factors of Eq.~\eqref{features:B:analytical} come from this term, the most interesting of which include: $(D_{22}-D_{11})^{n-1}$, which implies that each rung of the ladder structure is dependent on the difference between the permanent dipoles of each state; $(E_1-E_2)^{-n}$, indicating that the signal strength increases with decreasing difference between the energies the electronic states, which accentuates the underlying role of state mixing; exponentials $\exp{\left[-(T_\text{NIR}^2/16n)(\omega-n\omega_\text{NIR})^2\right]}$ which set the locations of the rungs of the ladder, as Gaussians centered at frequencies $\omega=n\, \omega_\text{NIR}$; and $F_{0,\text{NIR}}^n$ which for weak and moderately intense fields strongly moderates the signal with increasing $n$, counteracting the effect of the other factors, which ultimately is the reason why more rungs are not visible in the spectra. In conclusion, Eq.~\eqref{features:B:analytical} demonstrates that the adiabatic part of the ladder structure can be attributed to the mixing of field free states and that the signal strength is strongly dependent on the permanent dipoles and energies of the system, along with the intensity of the incoming NIR field. 

\subsubsection{Polar versus non-polar molecules}
The LISs and the ladder structures in the ATA spectra depend critically on the difference between the total permanent dipole moments of the electronic states $D_{22}-D_{11}$ [see Eqs.~\eqref{features:A:phase} and \eqref{features:B:analytical}]. Heteronuclear non-polar systems such as HD${}^+$ and HD would have a nonzero total permanent dipole moment due to the nuclear mass asymmetry, but since the electronic states are parity eigenstates, one would have $D_{11}=D_{22}$, and the aforementioned features would not be present. A necessary condition for our observations is therefore to consider polar molecules.

\subsection{\label{subsec:ovsa}Orientated and aligned targets}

All calculations up until this point have been made with the assumption of a fixed orientation of all molecules in the target with respect to the incoming field, as depicted in Fig.~\ref{fig:molecule}. Another pertinent arrangement of the target is alignment with respect to the incoming field, where all molecules are in one of two orientations, either as in Fig.~\ref{fig:molecule}, or opposite, i.e. with the two atoms interchanged. Both of these arrangements are experimentally feasible, with alignment~\cite{stapelfeldt2003} being simpler to realize than orientation~\cite{holmegaard2009,de2009,oda2010,frumker2012}. Here we compare the ATA spectra obtained under both of these circumstances.

Calculating the theoretical response from a system with a flipped orientation is equivalent to using a carrier-envelope phase $\varphi=\pi$ instead of $\varphi=0$ for both fields. Due to the nature of the response function in Eq.~\eqref{theory:A:response_1}, which features a linear factor of the dipole moment, the contributions from two calculations, with $\varphi=\pi$ and $\varphi=0$, can be added linearly, i.e. incoherently:
\begin{equation}
S_\text{aligned}(\omega,\tau)=\frac{1}{2}\left[S(\omega,\tau;\varphi=0) + S(\omega,\tau;\varphi=\pi)\right]. \label{results:C:response}
\end{equation}

\begin{figure*}
\includegraphics[width=0.64\textwidth]{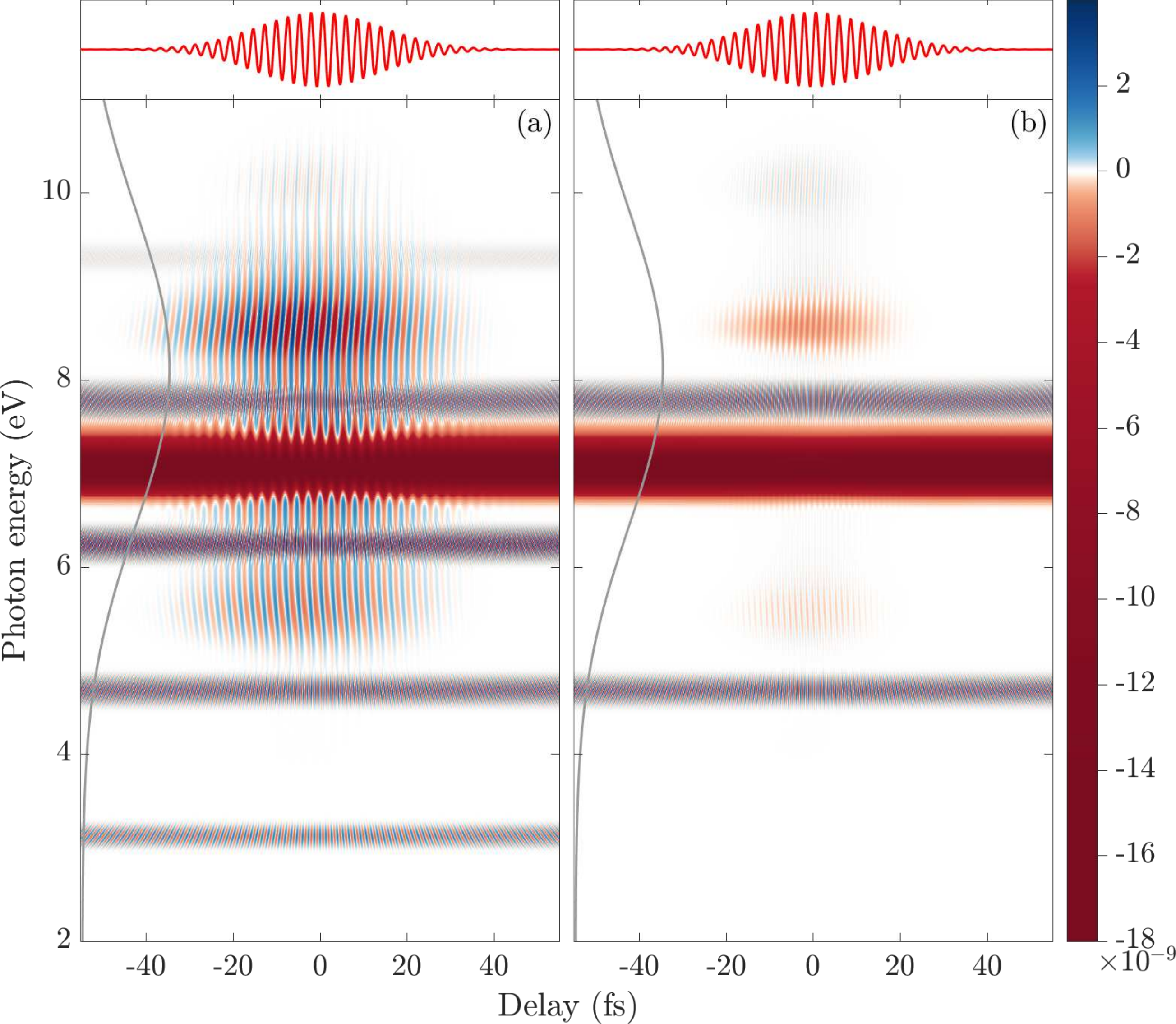}
\caption{\label{fig:ovsa} Attosecond transient absorption spectrum of the LiF molecule as calculated by Eq.~\eqref{theory:A:response_1}, exhibiting the difference between oriented [Panel~(a)] and aligned [Panel~(b)] targets. Both spectra were calculated using the full numerical method of Sec.~\ref{subsec:TDSE}. In panel~(a) where the carrier-envelope phase $\varphi$ of the fields is set to zero, corresponding to a target where all molecules are oriented relative to the field according to Fig.~\ref{fig:molecule}. Panel~(b) corresponds to Eq.~\eqref{results:C:response}, i.e. with a target where molecules are aligned with respect to the field. The frequency bandwidth of the UV pulse is shown in gray in both figures. The top panels depict the NIR pulse centered at $\tau=0$~fs. The color scale on the right displays the signal strength in arbitrary units. The pulse parameters are given in the text following Eq.~\eqref{theory:A:pulse}.} 
\end{figure*} 

Figure~\ref{fig:ovsa} demonstrates the difference in the spectrum when the target consists of oriented molecules [Fig.~\ref{fig:ovsa}~(a)], compared to when the target consists of aligned molecules [Fig.~\ref{fig:ovsa}~(b)]. Both spectra are calculated by the full numerical model of Sec.~\ref{subsec:TDSE}. Three distinctions are conspicuous; first, the LISs are significantly suppressed when using a target of aligned molecules. Second, the interference pattern visible in the absorption line in Fig.~\ref{fig:ovsa}~(a) is extinguished in Fig.~\ref{fig:ovsa}~(b). This pattern was attributed to which-way interference, due to more than one process in which population is transferred to the same final state, here either via the absorption of one UV photon or via the absorption of one UV photon and the subsequent absorption of one NIR photon or emission of one or two NIR photons. This is therefore connected with the first point, since the suppression of LISs suggests that several of the interfering pathways are unavailable. However, an important caveat is that we do not argue that the processes themselves do not occur in the molecules of an aligned target, but that due to opposite signs in the two spectra corresponding to opposite orientations, they will add destructively and not appear in $S_\text{aligned}$. The third distinction between the spectra is that, unlike in Fig.~\ref{fig:ovsa}~(a), in Fig.~\ref{fig:ovsa}~(b) only rungs corresponding to odd-multiples of the NIR photon energy are visible. The cause of the vanishing rungs can be understood by considering Eq.~\eqref{features:B:analytical}, and presuming that the following reasoning can be extended also to the non-adiabatic part of the ladder. If $\varphi=0$, we have $\exp{[i(n-1)\varphi]}=1$; if $\varphi=\pi$, then $\exp{[i(n-1)\varphi]}=(-1)^{n-1}$. Hence, the two terms in Eq.~\eqref{results:C:response} will have opposite signs if $n$ is even, and they will cancel each other out. 

To summarize, the ATAS results for a system with a permanent dipole can be expected to differ significantly depending on the arrangement of the target. From a collection of oriented molecules we can expect a richer spectrogram to emerge, relative to a collection of aligned molecules with two opposite orientations, as in the latter case some features will be suppressed due to contributions carrying opposite signs. 

\section{\label{section:conclusion}Conclusion and outlook}

In conclusion, we have presented, to our knowledge, the first application and theoretical investigation of ATAS on oriented or aligned polar molecules. To illustrate the characteristic effects induced by the presence of a permanent dipole, we used the polar diatomic molecule LiF as an example. We found two novel structures in the ATA spectra of polar molecules that are absent in ATA spectra of atoms and homonuclear molecules. Firstly, since the electronic eigenstates have no definite parity, two-photon transitions to the excited state can occur, which leads to LISs located around the main absorption line; this is in contrast to systems lacking a permanent dipole, where LISs are only observed around energies corresponding to dark states. Secondly, the permanent dipole moments induce a ladder structure separated by the NIR photon energies in the ATA spectra. We characterized these structures by presenting a model with fixed nuclei in the adiabatic time-dependent basis. Analytical and semi-analytical expressions were derived for the LISs and ladder structures, which showed the origin of these structures as well as their parameter-dependencies. For example the presence of the structures were shown to depend on $D_{11}-D_{22}$, which implies that the atoms and heteronuclear non-polar molecules such as HD$^+$ will not exhibit the aforementioned structures.

Our current study supplements the already vast preexisting knowledge of ATAS in atoms and homonuclear molecules by addressing effects in ATAS of a nonzero permanent dipole. Since aligned molecular samples are experimentally easier to achieve than orientated molecular samples, we have further studied how the ATA spectra could look like for a aligned sample. Thus we believe that the necessary experimental capabilities are already in place for ATAS studies in polar systems.

\section*{Acknowledgments}

This research was supported by the Villum Kann Rasmussen center of excellence, QUSCOPE - Quantum Scale Optical Processes, and the Danish Council for Independent Research (Grant no. 7014-00092B). The numerical results were obtained at the Centre for Scientific Computing Aarhus (CSCAA). We thank Brett Esry and Greg Armstrong for providing tabulated energies and dipole couplings for the LiF molecule. 

\appendix

\section{\label{section:appendix_UV}Fourier transform of $\boldsymbol{\mathit{F}_\text{UV}}$} 
In this Appendix, we derive the analytical expression for the UV field in the frequency domain, $\tilde{F}_\text{UV}(\omega,\tau)$, used in the derivations in Sec.~\ref{subsubsec:ladder}. Starting with Eq.~\eqref{theory:A:pulse}, setting $t'=t-t_c$, we obtain
\begin{equation}
\begin{aligned}
F_\text{UV}(t')=&\,\frac{8F_{0,\text{UV}}}{\omega_\text{UV}T_\text{UV}^2}t'e^{-\frac{4t'^2}{T_\text{UV}^2}}\cos{(\omega_\text{UV} t'+\varphi)} \\
&+ F_{0,\text{UV}}e^{-\frac{4t'^2}{T_\text{UV}^2}}\sin{(\omega_\text{UV} t'+\varphi)}\label{app:A:initial}\\
=&\,\frac{4F_{0,\text{UV}}}{\omega_\text{UV}T_\text{UV}^2}t'e^{-\frac{4t'^2}{T_\text{UV}^2}}e^{i\omega_\text{UV}t'}e^{i\varphi} \\
&+ \frac{F_{0,\text{UV}}}{2i}e^{-\frac{4t'^2}{T_\text{UV}^2}}e^{i\omega_\text{UV}t'}e^{i\varphi},
\end{aligned}
\end{equation}
where we have neglected terms containing $e^{-i\omega_\text{UV}t'}$, as they correspond to shifts to negative frequencies. To proceed with the Fourier transform of Eq.~\eqref{app:A:initial}, it is necessary to note a number of relations \cite{kammler2007} that will be applied. Under the convention $\mathcal{F}[f(t)]=\tilde{f}(\omega)$, we have
\begin{align}
\mathcal{F}[e^{-\beta t^2}]&=\frac{1}{\sqrt{2\beta}}e^{-\frac{\omega^2}{4\beta}}\label{app:A:rel1}\\
\mathcal{F}[tf(t)]&=i\frac{d\tilde{f}(\omega)}{d\omega}\label{app:A:rel2}\\
\mathcal{F}[e^{ibt}f(t)]&=\tilde{f}(\omega-b)\label{app:A:rel3}\\
\mathcal{F}[f(t-\tau)]&=e^{-i\tau\omega}\hat{f}(\omega).\label{app:A:rel4}
\end{align}
Applying the relations of Eqs.~\eqref{app:A:rel1}-\eqref{app:A:rel4} in the Fourier transform of Eq.~\eqref{app:A:initial} yields the following expression for the UV pulse in the frequency domain
\begin{equation}
\tilde{F}_\text{UV}(\omega,\tau)= -i\frac{F_{0,\text{UV}}T_{UV}}{4\sqrt{2}}\frac{\omega}{\omega_\text{UV}}e^{-i\tau\omega}e^{i\varphi}e^{-\frac{T_\text{UV}^2}{16}(\omega-\omega_\text{UV})^2}. \label{app:A:final}
\end{equation}

\section{\label{section:appendix_IR}Fourier transform of $\boldsymbol{\mathit{F}_\text{NIR}^n}$}
In this Appendix, we calculate the Fourier transform of the $n$'th power of the NIR field $F_\text{NIR}^n$, where $n=\{2,3,4\}$, which is used in the derivations in Sec.~\ref{subsubsec:ladder}. Starting from Eq.~\eqref{theory:A:pulse}, we keep in mind that the NIR pulse is centered at $t_c=0$. The NIR field has a long period $T_\text{NIR}$ relative to the UV field, implying that the term containing a factor $T_\text{NIR}^{-2}$ is negligible, and so the field to the $n$'th power can be expressed as
\begin{equation}
\begin{aligned}
F^n_\text{NIR} &= F^n_{0,\text{NIR}}e^{-\frac{4nt^2}{T_\text{NIR}^2}}\sin^n(\omega_\text{NIR}t+\varphi)\\
&=F^n_{0,\text{NIR}}e^{-\frac{4nt^2}{T_\text{NIR}^2}}\frac{1}{(2i)^n}\left(e^{i\omega_\text{NIR}t}e^{i\varphi}-e^{-i\omega_\text{NIR}t}e^{-i\varphi}\right)^{\!n}.
\end{aligned}
\end{equation}

The term corresponding to the envelope of the pulse, $\exp{(-4nt^2/T^2_{\text{NIR}})}$, can be transformed according to Eq.~\eqref{app:A:rel1}, whereas the exponentials corresponding to the carrier factor, $\exp{(in\omega_\text{NIR}t)}$, translate to shifts in energy according to Eq.~\eqref{app:A:rel4}. Considering that positive exponents imply shifts to positive frequencies, we keep only the terms of the expanded factor $(e^{i\omega_\text{NIR}t}-e^{-i\omega_\text{NIR}t})^{n}$ that are equal to $\exp{(in\omega_\text{NIR}t)}$ with $n=\{2,3,4\}$. The expansion of the parenthesis for $n=4$ yield two relevant terms, $\exp{(i2\omega_\text{NIR}t)}$ and $\exp{(i4\omega_\text{NIR}t)}$, but the contribution from the former will be small compared to the contribution from the equivalent term in the $n=2$ expansion, and so we can neglect it. Under these circumstances, the general expression is simply
\begin{equation}
F^n_\text{NIR} =F^n_{0,\text{NIR}}e^{-\frac{4nt^2}{T_\text{NIR}^2}}\frac{1}{(2i)^n}e^{in\omega_\text{NIR}t}e^{in\varphi},
\end{equation}
which upon Fourier transform yields
\begin{equation}
\begin{aligned}
\mathcal{F}\left[F_\text{NIR}^n\right]=&\,\left(-\frac{i}{2}\right)^{n}F^n_{0,\text{NIR}} \frac{T_\text{NIR}}{2\sqrt{2n}}\\
&\times e^{in\varphi}e^{-\frac{T_\text{NIR}^2}{16n}(\omega-n\omega_\text{NIR})^2}. \label{app:B:final}
\end{aligned}
\end{equation}


\begin{thebibliography}{49}%
\makeatletter
\providecommand \@ifxundefined [1]{%
 \@ifx{#1\undefined}
}%
\providecommand \@ifnum [1]{%
 \ifnum #1\expandafter \@firstoftwo
 \else \expandafter \@secondoftwo
 \fi
}%
\providecommand \@ifx [1]{%
 \ifx #1\expandafter \@firstoftwo
 \else \expandafter \@secondoftwo
 \fi
}%
\providecommand \natexlab [1]{#1}%
\providecommand \enquote  [1]{``#1''}%
\providecommand \bibnamefont  [1]{#1}%
\providecommand \bibfnamefont [1]{#1}%
\providecommand \citenamefont [1]{#1}%
\providecommand \href@noop [0]{\@secondoftwo}%
\providecommand \href [0]{\begingroup \@sanitize@url \@href}%
\providecommand \@href[1]{\@@startlink{#1}\@@href}%
\providecommand \@@href[1]{\endgroup#1\@@endlink}%
\providecommand \@sanitize@url [0]{\catcode `\\12\catcode `\$12\catcode
  `\&12\catcode `\#12\catcode `\^12\catcode `\_12\catcode `\%12\relax}%
\providecommand \@@startlink[1]{}%
\providecommand \@@endlink[0]{}%
\providecommand \url  [0]{\begingroup\@sanitize@url \@url }%
\providecommand \@url [1]{\endgroup\@href {#1}{\urlprefix }}%
\providecommand \urlprefix  [0]{URL }%
\providecommand \Eprint [0]{\href }%
\providecommand \doibase [0]{http://dx.doi.org/}%
\providecommand \selectlanguage [0]{\@gobble}%
\providecommand \bibinfo  [0]{\@secondoftwo}%
\providecommand \bibfield  [0]{\@secondoftwo}%
\providecommand \translation [1]{[#1]}%
\providecommand \BibitemOpen [0]{}%
\providecommand \bibitemStop [0]{}%
\providecommand \bibitemNoStop [0]{.\EOS\space}%
\providecommand \EOS [0]{\spacefactor3000\relax}%
\providecommand \BibitemShut  [1]{\csname bibitem#1\endcsname}%
\let\auto@bib@innerbib\@empty
\bibitem [{\citenamefont {Bengtsson}\ \emph {et~al.}(2017)\citenamefont
  {Bengtsson}, \citenamefont {Larsen}, \citenamefont {Kroon}, \citenamefont
  {Camp}, \citenamefont {Miranda}, \citenamefont {Arnold}, \citenamefont
  {L'Huillier}, \citenamefont {Schafer}, \citenamefont {Gaarde}, \citenamefont
  {Rippe} \emph {et~al.}}]{bengtsson2017}%
  \BibitemOpen
  \bibfield  {author} {\bibinfo {author} {\bibfnamefont {S.}~\bibnamefont
  {Bengtsson}}, \bibinfo {author} {\bibfnamefont {E.~W.}\ \bibnamefont
  {Larsen}}, \bibinfo {author} {\bibfnamefont {D.}~\bibnamefont {Kroon}},
  \bibinfo {author} {\bibfnamefont {S.}~\bibnamefont {Camp}}, \bibinfo {author}
  {\bibfnamefont {M.}~\bibnamefont {Miranda}}, \bibinfo {author} {\bibfnamefont
  {C.~L.}\ \bibnamefont {Arnold}}, \bibinfo {author} {\bibfnamefont
  {A.}~\bibnamefont {L'Huillier}}, \bibinfo {author} {\bibfnamefont {K.~J.}\
  \bibnamefont {Schafer}}, \bibinfo {author} {\bibfnamefont {M.~B.}\
  \bibnamefont {Gaarde}}, \bibinfo {author} {\bibfnamefont {L.}~\bibnamefont
  {Rippe}},  \emph {et~al.},\ }\bibfield  {title} {\enquote {\bibinfo {title}
  {Space--time control of free induction decay in the extreme ultraviolet},}\
  }\href@noop {} {\bibfield  {journal} {\bibinfo  {journal} {Nature Photonics}\
  }\textbf {\bibinfo {volume} {11}},\ \bibinfo {pages} {252} (\bibinfo {year}
  {2017})}\BibitemShut {NoStop}%
\bibitem [{\citenamefont {Calegari}\ \emph {et~al.}(2016)\citenamefont
  {Calegari}, \citenamefont {Sansone}, \citenamefont {Stagira}, \citenamefont
  {Vozzi},\ and\ \citenamefont {Nisoli}}]{calegari2016advances}%
  \BibitemOpen
  \bibfield  {author} {\bibinfo {author} {\bibfnamefont {F.}~\bibnamefont
  {Calegari}}, \bibinfo {author} {\bibfnamefont {G.}~\bibnamefont {Sansone}},
  \bibinfo {author} {\bibfnamefont {S.}~\bibnamefont {Stagira}}, \bibinfo
  {author} {\bibfnamefont {C.}~\bibnamefont {Vozzi}}, \ and\ \bibinfo {author}
  {\bibfnamefont {M.}~\bibnamefont {Nisoli}},\ }\bibfield  {title} {\enquote
  {\bibinfo {title} {Advances in attosecond science},}\ }\href@noop {}
  {\bibfield  {journal} {\bibinfo  {journal} {J. Phys. B}\ }\textbf {\bibinfo {volume} {49}},\ \bibinfo
  {pages} {062001} (\bibinfo {year} {2016})}\BibitemShut {NoStop}%
\bibitem [{\citenamefont {Loh}\ \emph {et~al.}(2008)\citenamefont {Loh},
  \citenamefont {Greene},\ and\ \citenamefont {R.}}]{loh2008}%
  \BibitemOpen
  \bibfield  {author} {\bibinfo {author} {\bibfnamefont {Z.-H.}\ \bibnamefont
  {Loh}}, \bibinfo {author} {\bibfnamefont {C.~H.}\ \bibnamefont {Greene}}, \
  and\ \bibinfo {author} {\bibfnamefont {Leone~S.}\ \bibnamefont {R.}},\
  }\bibfield  {title} {\enquote {\bibinfo {title} {Femtosecond induced
  transparency and absorption in the extreme ultraviolet by coherent coupling
  of the he 2s2p (${}^1$P${}^\text{o}$) and 2p2 (${}^\text{1}$S${}^\text{e}$) double excitation states with 800~nm
  light},}\ }\href {\doibase https://doi.org/10.1016/j.chemphys.2007.11.005}
  {\bibfield  {journal} {\bibinfo  {journal} {Chemical Physics}\ }\textbf
  {\bibinfo {volume} {350}},\ \bibinfo {pages} {7 -- 13} (\bibinfo {year}
  {2008})}\BibitemShut {NoStop}%
\bibitem [{\citenamefont {Goulielmakis}\ \emph {et~al.}(2010)\citenamefont
  {Goulielmakis}, \citenamefont {Loh}, \citenamefont {Wirth}, \citenamefont
  {Santra}, \citenamefont {Rohringer}, \citenamefont {Yakovlev}, \citenamefont
  {Zherebtsov}, \citenamefont {Pfeifer}, \citenamefont {Azzeer}, \citenamefont
  {Kling} \emph {et~al.}}]{goulielmakis2010}%
  \BibitemOpen
  \bibfield  {author} {\bibinfo {author} {\bibfnamefont {E.}~\bibnamefont
  {Goulielmakis}}, \bibinfo {author} {\bibfnamefont {Z.-H.}\ \bibnamefont
  {Loh}}, \bibinfo {author} {\bibfnamefont {A.}~\bibnamefont {Wirth}}, \bibinfo
  {author} {\bibfnamefont {R.}~\bibnamefont {Santra}}, \bibinfo {author}
  {\bibfnamefont {N.}~\bibnamefont {Rohringer}}, \bibinfo {author}
  {\bibfnamefont {V.~S.}\ \bibnamefont {Yakovlev}}, \bibinfo {author}
  {\bibfnamefont {S.}~\bibnamefont {Zherebtsov}}, \bibinfo {author}
  {\bibfnamefont {T.}~\bibnamefont {Pfeifer}}, \bibinfo {author} {\bibfnamefont
  {A.~M.}\ \bibnamefont {Azzeer}}, \bibinfo {author} {\bibfnamefont {M.~F.}\
  \bibnamefont {Kling}},  \emph {et~al.},\ }\bibfield  {title} {\enquote
  {\bibinfo {title} {Real-time observation of valence electron motion},}\
  }\href@noop {} {\bibfield  {journal} {\bibinfo  {journal} {Nature}\ }\textbf
  {\bibinfo {volume} {466}},\ \bibinfo {pages} {739} (\bibinfo {year}
  {2010})}\BibitemShut {NoStop}%
\bibitem [{\citenamefont {Beck}\ \emph {et~al.}(2015)\citenamefont {Beck},
  \citenamefont {Neumark},\ and\ \citenamefont {Leone}}]{beck2015probing}%
  \BibitemOpen
  \bibfield  {author} {\bibinfo {author} {\bibfnamefont {A.~R.}\ \bibnamefont
  {Beck}}, \bibinfo {author} {\bibfnamefont {D.~M.}\ \bibnamefont {Neumark}}, \
  and\ \bibinfo {author} {\bibfnamefont {S.~R.}\ \bibnamefont {Leone}},\
  }\bibfield  {title} {\enquote {\bibinfo {title} {Probing ultrafast dynamics
  with attosecond transient absorption},}\ }\href@noop {} {\bibfield  {journal}
  {\bibinfo  {journal} {Chem. Phys. Lett.}\ }\textbf {\bibinfo {volume}
  {624}},\ \bibinfo {pages} {119--130} (\bibinfo {year} {2015})}\BibitemShut
  {NoStop}%
\bibitem [{\citenamefont {Wu}\ \emph {et~al.}(2016)\citenamefont {Wu},
  \citenamefont {Chen}, \citenamefont {Camp}, \citenamefont {Schafer},\ and\
  \citenamefont {Gaarde}}]{wu2016theory}%
  \BibitemOpen
  \bibfield  {author} {\bibinfo {author} {\bibfnamefont {M.}~\bibnamefont
  {Wu}}, \bibinfo {author} {\bibfnamefont {S.}~\bibnamefont {Chen}}, \bibinfo
  {author} {\bibfnamefont {S.}~\bibnamefont {Camp}}, \bibinfo {author}
  {\bibfnamefont {K.~J.}\ \bibnamefont {Schafer}}, \ and\ \bibinfo {author}
  {\bibfnamefont {M.~B.}\ \bibnamefont {Gaarde}},\ }\bibfield  {title}
  {\enquote {\bibinfo {title} {Theory of strong-field attosecond transient
  absorption},}\ }\href@noop {} {\bibfield  {journal} {\bibinfo  {journal}
  {J. Phys. B}\ }\textbf
  {\bibinfo {volume} {49}},\ \bibinfo {pages} {062003} (\bibinfo {year}
  {2016})}\BibitemShut {NoStop}%
\bibitem [{\citenamefont {Wang}\ \emph {et~al.}(2010)\citenamefont {Wang},
  \citenamefont {Chini}, \citenamefont {Chen}, \citenamefont {Zhang},
  \citenamefont {He}, \citenamefont {Cheng}, \citenamefont {Wu}, \citenamefont
  {Thumm},\ and\ \citenamefont {Chang}}]{wang2010attosecond}%
  \BibitemOpen
  \bibfield  {author} {\bibinfo {author} {\bibfnamefont {H.}~\bibnamefont
  {Wang}}, \bibinfo {author} {\bibfnamefont {M.}~\bibnamefont {Chini}},
  \bibinfo {author} {\bibfnamefont {S.}~\bibnamefont {Chen}}, \bibinfo {author}
  {\bibfnamefont {C.-H.}\ \bibnamefont {Zhang}}, \bibinfo {author}
  {\bibfnamefont {F.}~\bibnamefont {He}}, \bibinfo {author} {\bibfnamefont
  {Y.}~\bibnamefont {Cheng}}, \bibinfo {author} {\bibfnamefont
  {Y.}~\bibnamefont {Wu}}, \bibinfo {author} {\bibfnamefont {U.}~\bibnamefont
  {Thumm}}, \ and\ \bibinfo {author} {\bibfnamefont {Z.}~\bibnamefont
  {Chang}},\ }\bibfield  {title} {\enquote {\bibinfo {title} {Attosecond
  time-resolved autoionization of argon},}\ }\href@noop {} {\bibfield
  {journal} {\bibinfo  {journal} {Phys. Rev. Lett.}\ }\textbf {\bibinfo
  {volume} {105}},\ \bibinfo {pages} {143002} (\bibinfo {year}
  {2010})}\BibitemShut {NoStop}%
\bibitem [{\citenamefont {Wirth}\ \emph {et~al.}(2011)\citenamefont {Wirth},
  \citenamefont {Hassan}, \citenamefont {Grgura{\v s}}, \citenamefont {Gagnon},
  \citenamefont {Moulet}, \citenamefont {Luu}, \citenamefont {Pabst},
  \citenamefont {Santra}, \citenamefont {Alahmed}, \citenamefont {Azzeer},
  \citenamefont {Yakovlev}, \citenamefont {Pervak}, \citenamefont {Krausz},\
  and\ \citenamefont {Goulielmakis}}]{wirth2011}%
  \BibitemOpen
  \bibfield  {author} {\bibinfo {author} {\bibfnamefont {A.}~\bibnamefont
  {Wirth}}, \bibinfo {author} {\bibfnamefont {M.~Th.}\ \bibnamefont {Hassan}},
  \bibinfo {author} {\bibfnamefont {I.}~\bibnamefont {Grgura{\v s}}}, \bibinfo
  {author} {\bibfnamefont {J.}~\bibnamefont {Gagnon}}, \bibinfo {author}
  {\bibfnamefont {A.}~\bibnamefont {Moulet}}, \bibinfo {author} {\bibfnamefont
  {T.~T.}\ \bibnamefont {Luu}}, \bibinfo {author} {\bibfnamefont
  {S.}~\bibnamefont {Pabst}}, \bibinfo {author} {\bibfnamefont
  {R.}~\bibnamefont {Santra}}, \bibinfo {author} {\bibfnamefont {Z.~A.}\
  \bibnamefont {Alahmed}}, \bibinfo {author} {\bibfnamefont {A.~M.}\
  \bibnamefont {Azzeer}}, \bibinfo {author} {\bibfnamefont {V.~S.}\
  \bibnamefont {Yakovlev}}, \bibinfo {author} {\bibfnamefont {V.}~\bibnamefont
  {Pervak}}, \bibinfo {author} {\bibfnamefont {F.}~\bibnamefont {Krausz}}, \
  and\ \bibinfo {author} {\bibfnamefont {E.}~\bibnamefont {Goulielmakis}},\
  }\bibfield  {title} {\enquote {\bibinfo {title} {Synthesized light
  transients},}\ }\href {\doibase 10.1126/science.1210268} {\bibfield
  {journal} {\bibinfo  {journal} {Science}\ }\textbf {\bibinfo {volume}
  {334}},\ \bibinfo {pages} {195--200} (\bibinfo {year} {2011})}\BibitemShut
  {NoStop}%
\bibitem [{\citenamefont {Ott}\ \emph {et~al.}(2014)\citenamefont {Ott},
  \citenamefont {Kaldun}, \citenamefont {Argenti}, \citenamefont {Raith},
  \citenamefont {Meyer}, \citenamefont {Laux}, \citenamefont {Zhang},
  \citenamefont {Bl{\"a}ttermann}, \citenamefont {Hagstotz}, \citenamefont
  {Ding} \emph {et~al.}}]{ott2014}%
  \BibitemOpen
  \bibfield  {author} {\bibinfo {author} {\bibfnamefont {C.}~\bibnamefont
  {Ott}}, \bibinfo {author} {\bibfnamefont {A.}~\bibnamefont {Kaldun}},
  \bibinfo {author} {\bibfnamefont {L.}~\bibnamefont {Argenti}}, \bibinfo
  {author} {\bibfnamefont {P.}~\bibnamefont {Raith}}, \bibinfo {author}
  {\bibfnamefont {K.}~\bibnamefont {Meyer}}, \bibinfo {author} {\bibfnamefont
  {M.}~\bibnamefont {Laux}}, \bibinfo {author} {\bibfnamefont {Y.}~\bibnamefont
  {Zhang}}, \bibinfo {author} {\bibfnamefont {A.}~\bibnamefont
  {Bl{\"a}ttermann}}, \bibinfo {author} {\bibfnamefont {S.}~\bibnamefont
  {Hagstotz}}, \bibinfo {author} {\bibfnamefont {T.}~\bibnamefont {Ding}},
  \emph {et~al.},\ }\bibfield  {title} {\enquote {\bibinfo {title}
  {Reconstruction and control of a time-dependent two-electron wave packet},}\
  }\href@noop {} {\bibfield  {journal} {\bibinfo  {journal} {Nature}\ }\textbf
  {\bibinfo {volume} {516}},\ \bibinfo {pages} {374--378} (\bibinfo {year}
  {2014})}\BibitemShut {NoStop}%
\bibitem [{\citenamefont {Holler}\ \emph {et~al.}(2011)\citenamefont {Holler},
  \citenamefont {Schapper}, \citenamefont {Gallmann},\ and\ \citenamefont
  {Keller}}]{holler2011}%
  \BibitemOpen
  \bibfield  {author} {\bibinfo {author} {\bibfnamefont {M.}~\bibnamefont
  {Holler}}, \bibinfo {author} {\bibfnamefont {F.}~\bibnamefont {Schapper}},
  \bibinfo {author} {\bibfnamefont {L.}~\bibnamefont {Gallmann}}, \ and\
  \bibinfo {author} {\bibfnamefont {U.}~\bibnamefont {Keller}},\ }\bibfield
  {title} {\enquote {\bibinfo {title} {Attosecond electron wave-packet
  interference observed by transient absorption},}\ }\href {\doibase
  10.1103/PhysRevLett.106.123601} {\bibfield  {journal} {\bibinfo  {journal}
  {Phys. Rev. Lett.}\ }\textbf {\bibinfo {volume} {106}},\ \bibinfo {pages}
  {123601} (\bibinfo {year} {2011})}\BibitemShut {NoStop}%
\bibitem [{\citenamefont {Pabst}\ \emph {et~al.}(2012)\citenamefont {Pabst},
  \citenamefont {Sytcheva}, \citenamefont {Moulet}, \citenamefont {Wirth},
  \citenamefont {Goulielmakis},\ and\ \citenamefont {Santra}}]{pabst2012}%
  \BibitemOpen
  \bibfield  {author} {\bibinfo {author} {\bibfnamefont {S.}~\bibnamefont
  {Pabst}}, \bibinfo {author} {\bibfnamefont {A.}~\bibnamefont {Sytcheva}},
  \bibinfo {author} {\bibfnamefont {A.}~\bibnamefont {Moulet}}, \bibinfo
  {author} {\bibfnamefont {A.}~\bibnamefont {Wirth}}, \bibinfo {author}
  {\bibfnamefont {E.}~\bibnamefont {Goulielmakis}}, \ and\ \bibinfo {author}
  {\bibfnamefont {R.}~\bibnamefont {Santra}},\ }\bibfield  {title} {\enquote
  {\bibinfo {title} {Theory of attosecond transient-absorption spectroscopy of
  krypton for overlapping pump and probe pulses},}\ }\href {\doibase
  10.1103/PhysRevA.86.063411} {\bibfield  {journal} {\bibinfo  {journal} {Phys.
  Rev. A}\ }\textbf {\bibinfo {volume} {86}},\ \bibinfo {pages} {063411}
  (\bibinfo {year} {2012})}\BibitemShut {NoStop}%
\bibitem [{\citenamefont {Kobayashi}\ \emph {et~al.}(2017)\citenamefont
  {Kobayashi}, \citenamefont {Timmers}, \citenamefont {Sabbar}, \citenamefont
  {Leone},\ and\ \citenamefont {Neumark}}]{kobayashi2017}%
  \BibitemOpen
  \bibfield  {author} {\bibinfo {author} {\bibfnamefont {Y.}~\bibnamefont
  {Kobayashi}}, \bibinfo {author} {\bibfnamefont {H.}~\bibnamefont {Timmers}},
  \bibinfo {author} {\bibfnamefont {M.}~\bibnamefont {Sabbar}}, \bibinfo
  {author} {\bibfnamefont {S.~R.}\ \bibnamefont {Leone}}, \ and\ \bibinfo
  {author} {\bibfnamefont {D.~M.}\ \bibnamefont {Neumark}},\ }\bibfield
  {title} {\enquote {\bibinfo {title} {Attosecond transient-absorption dynamics
  of xenon core-excited states in a strong driving field},}\ }\href {\doibase
  10.1103/PhysRevA.95.031401} {\bibfield  {journal} {\bibinfo  {journal} {Phys.
  Rev. A}\ }\textbf {\bibinfo {volume} {95}},\ \bibinfo {pages} {031401}
  (\bibinfo {year} {2017})}\BibitemShut {NoStop}%
\bibitem [{\citenamefont {Sabbar}\ \emph {et~al.}(2017)\citenamefont {Sabbar},
  \citenamefont {Timmers}, \citenamefont {Chen}, \citenamefont {Pymer},
  \citenamefont {Loh}, \citenamefont {Sayres}, \citenamefont {Pabst},
  \citenamefont {Santra},\ and\ \citenamefont {Leone}}]{sabbar2017}%
  \BibitemOpen
  \bibfield  {author} {\bibinfo {author} {\bibfnamefont {M.}~\bibnamefont
  {Sabbar}}, \bibinfo {author} {\bibfnamefont {H.}~\bibnamefont {Timmers}},
  \bibinfo {author} {\bibfnamefont {Y.-J.}\ \bibnamefont {Chen}}, \bibinfo
  {author} {\bibfnamefont {A.~K.}\ \bibnamefont {Pymer}}, \bibinfo {author}
  {\bibfnamefont {Z.-H.}\ \bibnamefont {Loh}}, \bibinfo {author} {\bibfnamefont
  {S.~G.}\ \bibnamefont {Sayres}}, \bibinfo {author} {\bibfnamefont
  {S.}~\bibnamefont {Pabst}}, \bibinfo {author} {\bibfnamefont
  {R.}~\bibnamefont {Santra}}, \ and\ \bibinfo {author} {\bibfnamefont {S.~R.}\
  \bibnamefont {Leone}},\ }\bibfield  {title} {\enquote {\bibinfo {title}
  {State-resolved attosecond reversible and irreversible dynamics in strong
  optical fields},}\ }\href@noop {} {\bibfield  {journal} {\bibinfo  {journal}
  {Nature Physics}\ } (\bibinfo {year} {2017})}\BibitemShut {NoStop}%
\bibitem [{\citenamefont {B\ae{}kh\o{}j}\ \emph {et~al.}(2015)\citenamefont
  {B\ae{}kh\o{}j}, \citenamefont {Yue},\ and\ \citenamefont
  {Madsen}}]{baekhoej2015}%
  \BibitemOpen
  \bibfield  {author} {\bibinfo {author} {\bibfnamefont {J.~E.}\ \bibnamefont
  {B\ae{}kh\o{}j}}, \bibinfo {author} {\bibfnamefont {L.}~\bibnamefont {Yue}},
  \ and\ \bibinfo {author} {\bibfnamefont {L.~B.}\ \bibnamefont {Madsen}},\
  }\bibfield  {title} {\enquote {\bibinfo {title} {Nuclear-motion effects in
  attosecond transient-absorption spectroscopy of molecules},}\ }\href
  {\doibase 10.1103/PhysRevA.91.043408} {\bibfield  {journal} {\bibinfo
  {journal} {Phys. Rev. A}\ }\textbf {\bibinfo {volume} {91}},\ \bibinfo
  {pages} {043408} (\bibinfo {year} {2015})}\BibitemShut {NoStop}%
\bibitem [{\citenamefont {Hollstein}\ \emph {et~al.}(2017)\citenamefont
  {Hollstein}, \citenamefont {Santra},\ and\ \citenamefont
  {Pfannkuche}}]{hollstein2017}%
  \BibitemOpen
  \bibfield  {author} {\bibinfo {author} {\bibfnamefont {M.}~\bibnamefont
  {Hollstein}}, \bibinfo {author} {\bibfnamefont {R.}~\bibnamefont {Santra}}, \
  and\ \bibinfo {author} {\bibfnamefont {D.}~\bibnamefont {Pfannkuche}},\
  }\bibfield  {title} {\enquote {\bibinfo {title} {Correlation-driven charge
  migration following double ionization and attosecond transient absorption
  spectroscopy},}\ }\href {\doibase 10.1103/PhysRevA.95.053411} {\bibfield
  {journal} {\bibinfo  {journal} {Phys. Rev. A}\ }\textbf {\bibinfo {volume}
  {95}},\ \bibinfo {pages} {053411} (\bibinfo {year} {2017})}\BibitemShut
  {NoStop}%
\bibitem [{\citenamefont {B\ae{}kh\o{}j}\ \emph {et~al.}(2018)\citenamefont
  {B\ae{}kh\o{}j}, \citenamefont {L\'ev\^eque},\ and\ \citenamefont
  {Madsen}}]{baekhoej2018}%
  \BibitemOpen
  \bibfield  {author} {\bibinfo {author} {\bibfnamefont {J.~E.}\ \bibnamefont
  {B\ae{}kh\o{}j}}, \bibinfo {author} {\bibfnamefont {C.}~\bibnamefont
  {L\'ev\^eque}}, \ and\ \bibinfo {author} {\bibfnamefont {L.~B.}\ \bibnamefont
  {Madsen}},\ }\bibfield  {title} {\enquote {\bibinfo {title} {Signatures of a
  conical intersection in attosecond transient absorption spectroscopy},}\
  }\href {\doibase 10.1103/PhysRevLett.121.023203} {\bibfield  {journal}
  {\bibinfo  {journal} {Phys. Rev. Lett.}\ }\textbf {\bibinfo {volume} {121}},\
  \bibinfo {pages} {023203} (\bibinfo {year} {2018})}\BibitemShut {NoStop}%
\bibitem [{\citenamefont {Cheng}\ \emph {et~al.}(2016)\citenamefont {Cheng},
  \citenamefont {Chini}, \citenamefont {Wang}, \citenamefont
  {Gonz\'alez-Castrillo}, \citenamefont {Palacios}, \citenamefont {Argenti},
  \citenamefont {Mart\'{\i}n},\ and\ \citenamefont {Chang}}]{cheng2016}%
  \BibitemOpen
  \bibfield  {author} {\bibinfo {author} {\bibfnamefont {Y.}~\bibnamefont
  {Cheng}}, \bibinfo {author} {\bibfnamefont {M.}~\bibnamefont {Chini}},
  \bibinfo {author} {\bibfnamefont {X.}~\bibnamefont {Wang}}, \bibinfo {author}
  {\bibfnamefont {A.}~\bibnamefont {Gonz\'alez-Castrillo}}, \bibinfo {author}
  {\bibfnamefont {A.}~\bibnamefont {Palacios}}, \bibinfo {author}
  {\bibfnamefont {L.}~\bibnamefont {Argenti}}, \bibinfo {author} {\bibfnamefont
  {F.}~\bibnamefont {Mart\'{\i}n}}, \ and\ \bibinfo {author} {\bibfnamefont
  {Z.}~\bibnamefont {Chang}},\ }\bibfield  {title} {\enquote {\bibinfo {title}
  {Reconstruction of an excited-state molecular wave packet with attosecond
  transient absorption spectroscopy},}\ }\href {\doibase
  10.1103/PhysRevA.94.023403} {\bibfield  {journal} {\bibinfo  {journal} {Phys.
  Rev. A}\ }\textbf {\bibinfo {volume} {94}},\ \bibinfo {pages} {023403}
  (\bibinfo {year} {2016})}\BibitemShut {NoStop}%
\bibitem [{\citenamefont {Warrick}\ \emph {et~al.}(2016)\citenamefont
  {Warrick}, \citenamefont {Cao}, \citenamefont {Neumark},\ and\ \citenamefont
  {Leone}}]{warrick2016}%
  \BibitemOpen
  \bibfield  {author} {\bibinfo {author} {\bibfnamefont {E.~R.}\ \bibnamefont
  {Warrick}}, \bibinfo {author} {\bibfnamefont {W.}~\bibnamefont {Cao}},
  \bibinfo {author} {\bibfnamefont {D.~M.}\ \bibnamefont {Neumark}}, \ and\
  \bibinfo {author} {\bibfnamefont {S.~R.}\ \bibnamefont {Leone}},\ }\bibfield
  {title} {\enquote {\bibinfo {title} {Probing the dynamics of rydberg and
  valence states of molecular nitrogen with attosecond transient absorption
  spectroscopy},}\ }\href@noop {} {\bibfield  {journal} {\bibinfo  {journal}
  {Phys. Chem. A}\ }\textbf {\bibinfo {volume} {120}},\
  \bibinfo {pages} {3165--3174} (\bibinfo {year} {2016})},\ \bibinfo {note}
  {pMID: 26862883}\BibitemShut {NoStop}%
\bibitem [{\citenamefont {Warrick}\ \emph {et~al.}(2017)\citenamefont
  {Warrick}, \citenamefont {B\ae{}kh\o{}j}, \citenamefont {Cao}, \citenamefont
  {Fidler}, \citenamefont {Jensen}, \citenamefont {Madsen}, \citenamefont
  {Leone},\ and\ \citenamefont {Neumark}}]{warrick2017}%
  \BibitemOpen
  \bibfield  {author} {\bibinfo {author} {\bibfnamefont {E.~R.}\ \bibnamefont
  {Warrick}}, \bibinfo {author} {\bibfnamefont {J.~E.}\ \bibnamefont
  {B\ae{}kh\o{}j}}, \bibinfo {author} {\bibfnamefont {W.}~\bibnamefont {Cao}},
  \bibinfo {author} {\bibfnamefont {A.~P.}\ \bibnamefont {Fidler}}, \bibinfo
  {author} {\bibfnamefont {F.}~\bibnamefont {Jensen}}, \bibinfo {author}
  {\bibfnamefont {L.~B.}\ \bibnamefont {Madsen}}, \bibinfo {author}
  {\bibfnamefont {S.~R.}\ \bibnamefont {Leone}}, \ and\ \bibinfo {author}
  {\bibfnamefont {D.~M.}\ \bibnamefont {Neumark}},\ }\bibfield  {title}
  {\enquote {\bibinfo {title} {Attosecond transient absorption spectroscopy of
  molecular nitrogen: Vibrational coherences in the $b'$ ${}^1\Sigma^+_\text{u}$ state},}\ }\href
  {http://www.sciencedirect.com/science/article/pii/S0009261417301215}
  {\bibfield  {journal} {\bibinfo  {journal} {Chem. Phys. Lett.}\ }
  (\bibinfo {year} {2017})}\BibitemShut {NoStop}%
\bibitem [{\citenamefont {Liao}\ \emph {et~al.}(2017)\citenamefont {Liao},
  \citenamefont {Li}, \citenamefont {Haxton}, \citenamefont {Rescigno},
  \citenamefont {Lucchese}, \citenamefont {McCurdy},\ and\ \citenamefont
  {Sandhu}}]{liao2017}%
  \BibitemOpen
  \bibfield  {author} {\bibinfo {author} {\bibfnamefont {C.-T.}\ \bibnamefont
  {Liao}}, \bibinfo {author} {\bibfnamefont {X.}~\bibnamefont {Li}}, \bibinfo
  {author} {\bibfnamefont {D.~J.}\ \bibnamefont {Haxton}}, \bibinfo {author}
  {\bibfnamefont {T.~N.}\ \bibnamefont {Rescigno}}, \bibinfo {author}
  {\bibfnamefont {R.~R.}\ \bibnamefont {Lucchese}}, \bibinfo {author}
  {\bibfnamefont {C.~W.}\ \bibnamefont {McCurdy}}, \ and\ \bibinfo {author}
  {\bibfnamefont {A.}~\bibnamefont {Sandhu}},\ }\bibfield  {title} {\enquote
  {\bibinfo {title} {Probing autoionizing states of molecular oxygen with xuv
  transient absorption: Electronic-symmetry-dependent line shapes and
  laser-induced modifications},}\ }\href {\doibase 10.1103/PhysRevA.95.043427}
  {\bibfield  {journal} {\bibinfo  {journal} {Phys. Rev. A}\ }\textbf {\bibinfo
  {volume} {95}},\ \bibinfo {pages} {043427} (\bibinfo {year}
  {2017})}\BibitemShut {NoStop}%
\bibitem [{\citenamefont {Schultze}\ \emph {et~al.}(2014)\citenamefont
  {Schultze}, \citenamefont {Ramasesha}, \citenamefont {Pemmaraju},
  \citenamefont {Sato}, \citenamefont {Whitmore}, \citenamefont {Gandman},
  \citenamefont {Prell}, \citenamefont {Borja}, \citenamefont {Prendergast},
  \citenamefont {Yabana}, \citenamefont {Neumark},\ and\ \citenamefont
  {Leone}}]{schultze2014}%
  \BibitemOpen
  \bibfield  {author} {\bibinfo {author} {\bibfnamefont {M.}~\bibnamefont
  {Schultze}}, \bibinfo {author} {\bibfnamefont {K.}~\bibnamefont {Ramasesha}},
  \bibinfo {author} {\bibfnamefont {C.~D.}\ \bibnamefont {Pemmaraju}}, \bibinfo
  {author} {\bibfnamefont {S.~A.}\ \bibnamefont {Sato}}, \bibinfo {author}
  {\bibfnamefont {D.}~\bibnamefont {Whitmore}}, \bibinfo {author}
  {\bibfnamefont {A.}~\bibnamefont {Gandman}}, \bibinfo {author} {\bibfnamefont
  {J.~S.}\ \bibnamefont {Prell}}, \bibinfo {author} {\bibfnamefont {L.~J.}\
  \bibnamefont {Borja}}, \bibinfo {author} {\bibfnamefont {D.}~\bibnamefont
  {Prendergast}}, \bibinfo {author} {\bibfnamefont {K.}~\bibnamefont {Yabana}},
  \bibinfo {author} {\bibfnamefont {D.~M.}\ \bibnamefont {Neumark}}, \ and\
  \bibinfo {author} {\bibfnamefont {Stephen~R.}\ \bibnamefont {Leone}},\
  }\bibfield  {title} {\enquote {\bibinfo {title} {Attosecond band-gap dynamics
  in silicon},}\ }\href {\doibase 10.1126/science.1260311} {\bibfield
  {journal} {\bibinfo  {journal} {Science}\ }\textbf {\bibinfo {volume}
  {346}},\ \bibinfo {pages} {1348--1352} (\bibinfo {year} {2014})}\BibitemShut
  {NoStop}%
\bibitem [{\citenamefont {Borja}\ \emph {et~al.}(2016)\citenamefont {Borja},
  \citenamefont {Z\"{u}rch}, \citenamefont {Pemmaraju}, \citenamefont
  {Schultze}, \citenamefont {Ramasesha}, \citenamefont {Gandman}, \citenamefont
  {Prell}, \citenamefont {D.}, \citenamefont {M.},\ and\ \citenamefont
  {R.}}]{borja2016}%
  \BibitemOpen
  \bibfield  {author} {\bibinfo {author} {\bibfnamefont {L.~J.}\ \bibnamefont
  {Borja}}, \bibinfo {author} {\bibfnamefont {M.}~\bibnamefont {Z\"{u}rch}},
  \bibinfo {author} {\bibfnamefont {C.~D.}\ \bibnamefont {Pemmaraju}}, \bibinfo
  {author} {\bibfnamefont {M.}~\bibnamefont {Schultze}}, \bibinfo {author}
  {\bibfnamefont {K.}~\bibnamefont {Ramasesha}}, \bibinfo {author}
  {\bibfnamefont {A.}~\bibnamefont {Gandman}}, \bibinfo {author} {\bibfnamefont
  {J.~S.}\ \bibnamefont {Prell}}, \bibinfo {author} {\bibfnamefont
  {Prendergast}\ \bibnamefont {D.}}, \bibinfo {author} {\bibfnamefont
  {Neumark~D.}\ \bibnamefont {M.}}, \ and\ \bibinfo {author} {\bibfnamefont
  {Leone~S.}\ \bibnamefont {R.}},\ }\bibfield  {title} {\enquote {\bibinfo
  {title} {Extreme ultraviolet transient absorption of solids from femtosecond
  to attosecond timescales},}\ }\href {\doibase 10.1364/JOSAB.33.000C57}
  {\bibfield  {journal} {\bibinfo  {journal} {J. Opt. Soc. Am. B}\ }\textbf
  {\bibinfo {volume} {33}},\ \bibinfo {pages} {C57--C64} (\bibinfo {year}
  {2016})}\BibitemShut {NoStop}%
\bibitem [{\citenamefont {Moulet}\ \emph {et~al.}(2017)\citenamefont {Moulet},
  \citenamefont {Bertrand}, \citenamefont {Klostermann}, \citenamefont
  {Guggenmos}, \citenamefont {Karpowicz},\ and\ \citenamefont
  {Goulielmakis}}]{moulet2017}%
  \BibitemOpen
  \bibfield  {author} {\bibinfo {author} {\bibfnamefont {A.}~\bibnamefont
  {Moulet}}, \bibinfo {author} {\bibfnamefont {J.~B.}\ \bibnamefont
  {Bertrand}}, \bibinfo {author} {\bibfnamefont {T.}~\bibnamefont
  {Klostermann}}, \bibinfo {author} {\bibfnamefont {A.}~\bibnamefont
  {Guggenmos}}, \bibinfo {author} {\bibfnamefont {N.}~\bibnamefont
  {Karpowicz}}, \ and\ \bibinfo {author} {\bibfnamefont {E.}~\bibnamefont
  {Goulielmakis}},\ }\bibfield  {title} {\enquote {\bibinfo {title} {Soft x-ray
  excitonics},}\ }\href {\doibase 10.1126/science.aan4737} {\bibfield
  {journal} {\bibinfo  {journal} {Science}\ }\textbf {\bibinfo {volume}
  {357}},\ \bibinfo {pages} {1134--1138} (\bibinfo {year} {2017})}\BibitemShut
  {NoStop}%
\bibitem [{\citenamefont {Liao}\ \emph {et~al.}(2015)\citenamefont {Liao},
  \citenamefont {Sandhu}, \citenamefont {Camp}, \citenamefont {Schafer},\ and\
  \citenamefont {Gaarde}}]{liao2015}%
  \BibitemOpen
  \bibfield  {author} {\bibinfo {author} {\bibfnamefont {C.-T.}\ \bibnamefont
  {Liao}}, \bibinfo {author} {\bibfnamefont {A.}~\bibnamefont {Sandhu}},
  \bibinfo {author} {\bibfnamefont {S.}~\bibnamefont {Camp}}, \bibinfo {author}
  {\bibfnamefont {K.~J.}\ \bibnamefont {Schafer}}, \ and\ \bibinfo {author}
  {\bibfnamefont {M.~B.}\ \bibnamefont {Gaarde}},\ }\bibfield  {title}
  {\enquote {\bibinfo {title} {Beyond the single-atom response in absorption
  line shapes: Probing a dense, laser-dressed helium gas with attosecond pulse
  trains},}\ }\href {\doibase 10.1103/PhysRevLett.114.143002} {\bibfield
  {journal} {\bibinfo  {journal} {Phys. Rev. Lett.}\ }\textbf {\bibinfo
  {volume} {114}},\ \bibinfo {pages} {143002} (\bibinfo {year}
  {2015})}\BibitemShut {NoStop}%
\bibitem [{\citenamefont {Liao}\ \emph {et~al.}(2016)\citenamefont {Liao},
  \citenamefont {Sandhu}, \citenamefont {Camp}, \citenamefont {Schafer},\ and\
  \citenamefont {Gaarde}}]{liao2016}%
  \BibitemOpen
  \bibfield  {author} {\bibinfo {author} {\bibfnamefont {C.-T.}\ \bibnamefont
  {Liao}}, \bibinfo {author} {\bibfnamefont {A.}~\bibnamefont {Sandhu}},
  \bibinfo {author} {\bibfnamefont {S.}~\bibnamefont {Camp}}, \bibinfo {author}
  {\bibfnamefont {K.~J.}\ \bibnamefont {Schafer}}, \ and\ \bibinfo {author}
  {\bibfnamefont {M.~B.}\ \bibnamefont {Gaarde}},\ }\bibfield  {title}
  {\enquote {\bibinfo {title} {Attosecond transient absorption in dense gases:
  Exploring the interplay between resonant pulse propagation and laser-induced
  line-shape control},}\ }\href {\doibase 10.1103/PhysRevA.93.033405}
  {\bibfield  {journal} {\bibinfo  {journal} {Phys. Rev. A}\ }\textbf {\bibinfo
  {volume} {93}},\ \bibinfo {pages} {033405} (\bibinfo {year}
  {2016})}\BibitemShut {NoStop}%
\bibitem [{\citenamefont {Chen}\ \emph {et~al.}(2012)\citenamefont {Chen},
  \citenamefont {Bell}, \citenamefont {Beck}, \citenamefont {Mashiko},
  \citenamefont {Wu}, \citenamefont {Pfeiffer}, \citenamefont {Gaarde},
  \citenamefont {Neumark}, \citenamefont {Leone},\ and\ \citenamefont
  {Schafer}}]{chen2012}%
  \BibitemOpen
  \bibfield  {author} {\bibinfo {author} {\bibfnamefont {S.}~\bibnamefont
  {Chen}}, \bibinfo {author} {\bibfnamefont {M.~J.}\ \bibnamefont {Bell}},
  \bibinfo {author} {\bibfnamefont {A.~R.}\ \bibnamefont {Beck}}, \bibinfo
  {author} {\bibfnamefont {H.}~\bibnamefont {Mashiko}}, \bibinfo {author}
  {\bibfnamefont {M.}~\bibnamefont {Wu}}, \bibinfo {author} {\bibfnamefont
  {A.~N.}\ \bibnamefont {Pfeiffer}}, \bibinfo {author} {\bibfnamefont {M.~B.}\
  \bibnamefont {Gaarde}}, \bibinfo {author} {\bibfnamefont {D.~M.}\
  \bibnamefont {Neumark}}, \bibinfo {author} {\bibfnamefont {S.~R.}\
  \bibnamefont {Leone}}, \ and\ \bibinfo {author} {\bibfnamefont {K.~J.}\
  \bibnamefont {Schafer}},\ }\bibfield  {title} {\enquote {\bibinfo {title}
  {Light-induced states in attosecond transient absorption spectra of
  laser-dressed helium},}\ }\href {\doibase 10.1103/PhysRevA.86.063408}
  {\bibfield  {journal} {\bibinfo  {journal} {Phys. Rev. A}\ }\textbf {\bibinfo
  {volume} {86}},\ \bibinfo {pages} {063408} (\bibinfo {year}
  {2012})}\BibitemShut {NoStop}%
\bibitem [{\citenamefont {B\ae{}kh\o{}j}\ and\ \citenamefont
  {Madsen}(2015)}]{baekhoej2015_2}%
  \BibitemOpen
  \bibfield  {author} {\bibinfo {author} {\bibfnamefont {J.~E.}\ \bibnamefont
  {B\ae{}kh\o{}j}}\ and\ \bibinfo {author} {\bibfnamefont {L.~B.}\ \bibnamefont
  {Madsen}},\ }\bibfield  {title} {\enquote {\bibinfo {title} {Light-induced
  structures in attosecond transient-absorption spectroscopy of molecules},}\
  }\href {\doibase 10.1103/PhysRevA.92.023407} {\bibfield  {journal} {\bibinfo
  {journal} {Phys. Rev. A}\ }\textbf {\bibinfo {volume} {92}},\ \bibinfo
  {pages} {023407} (\bibinfo {year} {2015})}\BibitemShut {NoStop}%
\bibitem [{\citenamefont {Chen}\ \emph
  {et~al.}(2013{\natexlab{a}})\citenamefont {Chen}, \citenamefont {Wu},
  \citenamefont {Gaarde},\ and\ \citenamefont {Schafer}}]{chen2013}%
  \BibitemOpen
  \bibfield  {author} {\bibinfo {author} {\bibfnamefont {S.}~\bibnamefont
  {Chen}}, \bibinfo {author} {\bibfnamefont {M.}~\bibnamefont {Wu}}, \bibinfo
  {author} {\bibfnamefont {M.~B.}\ \bibnamefont {Gaarde}}, \ and\ \bibinfo
  {author} {\bibfnamefont {K.~J.}\ \bibnamefont {Schafer}},\ }\bibfield
  {title} {\enquote {\bibinfo {title} {Quantum interference in attosecond
  transient absorption of laser-dressed helium atoms},}\ }\href {\doibase
  10.1103/PhysRevA.87.033408} {\bibfield  {journal} {\bibinfo  {journal} {Phys.
  Rev. A}\ }\textbf {\bibinfo {volume} {87}},\ \bibinfo {pages} {033408}
  (\bibinfo {year} {2013}{\natexlab{a}})}\BibitemShut {NoStop}%
\bibitem [{\citenamefont {Chini}\ \emph {et~al.}(2014)\citenamefont {Chini},
  \citenamefont {Wang}, \citenamefont {Cheng},\ and\ \citenamefont
  {Chang}}]{chini2014}%
  \BibitemOpen
  \bibfield  {author} {\bibinfo {author} {\bibfnamefont {M.}~\bibnamefont
  {Chini}}, \bibinfo {author} {\bibfnamefont {X.}~\bibnamefont {Wang}},
  \bibinfo {author} {\bibfnamefont {Y.}~\bibnamefont {Cheng}}, \ and\ \bibinfo
  {author} {\bibfnamefont {Z.}~\bibnamefont {Chang}},\ }\bibfield  {title}
  {\enquote {\bibinfo {title} {Resonance effects and quantum beats in
  attosecond transient absorption of helium},}\ }\href
  {http://stacks.iop.org/0953-4075/47/i=12/a=124009} {\bibfield  {journal}
  {\bibinfo  {journal} {J. Phys. B}\ }\textbf {\bibinfo {volume} {47}},\ \bibinfo {pages} {124009}
  (\bibinfo {year} {2014})}\BibitemShut {NoStop}%
\bibitem [{\citenamefont {Lindberg}\ and\ \citenamefont
  {Koch}(1988)}]{lindberg1988}%
  \BibitemOpen
  \bibfield  {author} {\bibinfo {author} {\bibfnamefont {M.}~\bibnamefont
  {Lindberg}}\ and\ \bibinfo {author} {\bibfnamefont {S.~W.}\ \bibnamefont
  {Koch}},\ }\bibfield  {title} {\enquote {\bibinfo {title} {Transient
  oscillations and dynamic stark effect in semiconductors},}\ }\href {\doibase
  10.1103/PhysRevB.38.7607} {\bibfield  {journal} {\bibinfo  {journal} {Phys.
  Rev. B}\ }\textbf {\bibinfo {volume} {38}},\ \bibinfo {pages} {7607--7614}
  (\bibinfo {year} {1988})}\BibitemShut {NoStop}%
\bibitem [{\citenamefont {Brito-Cruz}\ \emph {et~al.}(1988)\citenamefont
  {Brito-Cruz}, \citenamefont {Gordon}, \citenamefont {Becker}, \citenamefont
  {Fork},\ and\ \citenamefont {Shank}}]{britocruz1988}%
  \BibitemOpen
  \bibfield  {author} {\bibinfo {author} {\bibfnamefont {C.~H.}\ \bibnamefont
  {Brito-Cruz}}, \bibinfo {author} {\bibfnamefont {J.~P.}\ \bibnamefont
  {Gordon}}, \bibinfo {author} {\bibfnamefont {P.~C.}\ \bibnamefont {Becker}},
  \bibinfo {author} {\bibfnamefont {R.~L.}\ \bibnamefont {Fork}}, \ and\
  \bibinfo {author} {\bibfnamefont {C.~V.}\ \bibnamefont {Shank}},\ }\bibfield
  {title} {\enquote {\bibinfo {title} {Dynamics of spectral hole burning},}\
  }\href {\doibase 10.1109/3.122} {\bibfield  {journal} {\bibinfo  {journal}
  {IEEE Journal of Quantum Electronics}\ }\textbf {\bibinfo {volume} {24}},\
  \bibinfo {pages} {261--269} (\bibinfo {year} {1988})}\BibitemShut {NoStop}%
\bibitem [{\citenamefont {Santra}\ \emph {et~al.}(2011)\citenamefont {Santra},
  \citenamefont {Yakovlev}, \citenamefont {Pfeifer},\ and\ \citenamefont
  {Loh}}]{santra2011}%
  \BibitemOpen
  \bibfield  {author} {\bibinfo {author} {\bibfnamefont {R.}~\bibnamefont
  {Santra}}, \bibinfo {author} {\bibfnamefont {V.~S.}\ \bibnamefont
  {Yakovlev}}, \bibinfo {author} {\bibfnamefont {T.}~\bibnamefont {Pfeifer}}, \
  and\ \bibinfo {author} {\bibfnamefont {Z.-H.}\ \bibnamefont {Loh}},\
  }\bibfield  {title} {\enquote {\bibinfo {title} {Theory of attosecond
  transient absorption spectroscopy of strong-field-generated ions},}\
  }\href@noop {} {\bibfield  {journal} {\bibinfo  {journal} {Phys. Rev.
  A}\ }\textbf {\bibinfo {volume} {83}},\ \bibinfo {pages} {033405} (\bibinfo
  {year} {2011})}\BibitemShut {NoStop}%
\bibitem [{\citenamefont {Pfeiffer}\ and\ \citenamefont
  {Leone}(2012)}]{pfeiffer2012}%
  \BibitemOpen
  \bibfield  {author} {\bibinfo {author} {\bibfnamefont {A.~N.}\ \bibnamefont
  {Pfeiffer}}\ and\ \bibinfo {author} {\bibfnamefont {S.~R.}\ \bibnamefont
  {Leone}},\ }\bibfield  {title} {\enquote {\bibinfo {title} {Transmission of
  an isolated attosecond pulse in a strong-field dressed atom},}\ }\href
  {\doibase 10.1103/PhysRevA.85.053422} {\bibfield  {journal} {\bibinfo
  {journal} {Phys. Rev. A}\ }\textbf {\bibinfo {volume} {85}},\ \bibinfo
  {pages} {053422} (\bibinfo {year} {2012})}\BibitemShut {NoStop}%
\bibitem [{\citenamefont {Chen}\ \emph
  {et~al.}(2013{\natexlab{b}})\citenamefont {Chen}, \citenamefont {Wu},
  \citenamefont {Gaarde},\ and\ \citenamefont {Schafer}}]{chen2013lip}%
  \BibitemOpen
  \bibfield  {author} {\bibinfo {author} {\bibfnamefont {S.}~\bibnamefont
  {Chen}}, \bibinfo {author} {\bibfnamefont {M.}~\bibnamefont {Wu}}, \bibinfo
  {author} {\bibfnamefont {M.~B.}\ \bibnamefont {Gaarde}}, \ and\ \bibinfo
  {author} {\bibfnamefont {K.~J.}\ \bibnamefont {Schafer}},\ }\bibfield
  {title} {\enquote {\bibinfo {title} {Laser-imposed phase in resonant
  absorption of an isolated attosecond pulse},}\ }\href {\doibase
  10.1103/PhysRevA.88.033409} {\bibfield  {journal} {\bibinfo  {journal} {Phys.
  Rev. A}\ }\textbf {\bibinfo {volume} {88}},\ \bibinfo {pages} {033409}
  (\bibinfo {year} {2013}{\natexlab{b}})}\BibitemShut {NoStop}%
\bibitem [{\citenamefont {Ott}\ \emph {et~al.}(2013)\citenamefont {Ott},
  \citenamefont {Kaldun}, \citenamefont {Raith}, \citenamefont {Meyer},
  \citenamefont {Laux}, \citenamefont {Evers}, \citenamefont {Keitel},
  \citenamefont {Greene},\ and\ \citenamefont {Pfeifer}}]{ott2013lorentz}%
  \BibitemOpen
  \bibfield  {author} {\bibinfo {author} {\bibfnamefont {C.}~\bibnamefont
  {Ott}}, \bibinfo {author} {\bibfnamefont {A.}~\bibnamefont {Kaldun}},
  \bibinfo {author} {\bibfnamefont {P.}~\bibnamefont {Raith}}, \bibinfo
  {author} {\bibfnamefont {K.}~\bibnamefont {Meyer}}, \bibinfo {author}
  {\bibfnamefont {M.}~\bibnamefont {Laux}}, \bibinfo {author} {\bibfnamefont
  {J.}~\bibnamefont {Evers}}, \bibinfo {author} {\bibfnamefont {C.~H.}\
  \bibnamefont {Keitel}}, \bibinfo {author} {\bibfnamefont {C.~H.}\
  \bibnamefont {Greene}}, \ and\ \bibinfo {author} {\bibfnamefont
  {T.}~\bibnamefont {Pfeifer}},\ }\bibfield  {title} {\enquote {\bibinfo
  {title} {Lorentz meets Fano in spectral line shapes: a universal phase and
  its laser control},}\ }\href@noop {} {\bibfield  {journal} {\bibinfo
  {journal} {Science}\ }\textbf {\bibinfo {volume} {340}},\ \bibinfo {pages}
  {716--720} (\bibinfo {year} {2013})}\BibitemShut {NoStop}%
\bibitem [{\citenamefont {Chini}\ \emph {et~al.}(2012)\citenamefont {Chini},
  \citenamefont {Zhao}, \citenamefont {Wang}, \citenamefont {Cheng},
  \citenamefont {Hu},\ and\ \citenamefont {Chang}}]{chini2012}%
  \BibitemOpen
  \bibfield  {author} {\bibinfo {author} {\bibfnamefont {M.}~\bibnamefont
  {Chini}}, \bibinfo {author} {\bibfnamefont {B.}~\bibnamefont {Zhao}},
  \bibinfo {author} {\bibfnamefont {H.}~\bibnamefont {Wang}}, \bibinfo {author}
  {\bibfnamefont {Y.}~\bibnamefont {Cheng}}, \bibinfo {author} {\bibfnamefont
  {S.~X.}\ \bibnamefont {Hu}}, \ and\ \bibinfo {author} {\bibfnamefont
  {Z.}~\bibnamefont {Chang}},\ }\bibfield  {title} {\enquote {\bibinfo {title}
  {Subcycle ac stark shift of helium excited states probed with isolated
  attosecond pulses},}\ }\href@noop {} {\bibfield  {journal} {\bibinfo
  {journal} {Phys. Rev. Lett.}\ }\textbf {\bibinfo {volume} {109}},\
  \bibinfo {pages} {073601} (\bibinfo {year} {2012})}\BibitemShut {NoStop}%
\bibitem [{\citenamefont {Autler}\ and\ \citenamefont
  {Townes}(1955)}]{autler1955}%
  \BibitemOpen
  \bibfield  {author} {\bibinfo {author} {\bibfnamefont {S.~H.}\ \bibnamefont
  {Autler}}\ and\ \bibinfo {author} {\bibfnamefont {C.~H.}\ \bibnamefont
  {Townes}},\ }\bibfield  {title} {\enquote {\bibinfo {title} {Stark effect in
  rapidly varying fields},}\ }\href {\doibase 10.1103/PhysRev.100.703}
  {\bibfield  {journal} {\bibinfo  {journal} {Phys. Rev.}\ }\textbf {\bibinfo
  {volume} {100}},\ \bibinfo {pages} {703--722} (\bibinfo {year}
  {1955})}\BibitemShut {NoStop}%
\bibitem [{\citenamefont {Wu}\ \emph {et~al.}(2013)\citenamefont {Wu},
  \citenamefont {Chen}, \citenamefont {Gaarde},\ and\ \citenamefont
  {Schafer}}]{wu2013}%
  \BibitemOpen
  \bibfield  {author} {\bibinfo {author} {\bibfnamefont {M.}~\bibnamefont
  {Wu}}, \bibinfo {author} {\bibfnamefont {S.}~\bibnamefont {Chen}}, \bibinfo
  {author} {\bibfnamefont {M.~B.}\ \bibnamefont {Gaarde}}, \ and\ \bibinfo
  {author} {\bibfnamefont {K.~J.}\ \bibnamefont {Schafer}},\ }\bibfield
  {title} {\enquote {\bibinfo {title} {Time-domain perspective on autler-townes
  splitting in attosecond transient absorption of laser-dressed helium
  atoms},}\ }\href {\doibase 10.1103/PhysRevA.88.043416} {\bibfield  {journal}
  {\bibinfo  {journal} {Phys. Rev. A}\ }\textbf {\bibinfo {volume} {88}},\
  \bibinfo {pages} {043416} (\bibinfo {year} {2013})}\BibitemShut {NoStop}%
\bibitem [{\citenamefont {R{\o}rstad}\ \emph {et~al.}(2017)\citenamefont
  {R{\o}rstad}, \citenamefont {B{\ae}kh{\o}j},\ and\ \citenamefont
  {Madsen}}]{rorstad2017}%
  \BibitemOpen
  \bibfield  {author} {\bibinfo {author} {\bibfnamefont {J.~J.}\ \bibnamefont
  {R{\o}rstad}}, \bibinfo {author} {\bibfnamefont {J.~E.}\ \bibnamefont
  {B{\ae}kh{\o}j}}, \ and\ \bibinfo {author} {\bibfnamefont {L.~B.}\
  \bibnamefont {Madsen}},\ }\bibfield  {title} {\enquote {\bibinfo {title}
  {Analytic modeling of structures in attosecond transient-absorption
  spectra},}\ }\href@noop {} {\bibfield  {journal} {\bibinfo  {journal}
  {Physical Review A}\ }\textbf {\bibinfo {volume} {96}},\ \bibinfo {pages}
  {013430} (\bibinfo {year} {2017})}\BibitemShut {NoStop}%
\bibitem [{\citenamefont {Gaarde}\ \emph {et~al.}(2011)\citenamefont {Gaarde},
  \citenamefont {Buth}, \citenamefont {Tate},\ and\ \citenamefont
  {Schafer}}]{gaarde2011}%
  \BibitemOpen
  \bibfield  {author} {\bibinfo {author} {\bibfnamefont {M.~B.}\ \bibnamefont
  {Gaarde}}, \bibinfo {author} {\bibfnamefont {C.}~\bibnamefont {Buth}},
  \bibinfo {author} {\bibfnamefont {J.~L.}\ \bibnamefont {Tate}}, \ and\
  \bibinfo {author} {\bibfnamefont {K.~J.}\ \bibnamefont {Schafer}},\
  }\bibfield  {title} {\enquote {\bibinfo {title} {Transient absorption and
  reshaping of ultrafast XUV light by laser-dressed helium},}\ }\href {\doibase
  10.1103/PhysRevA.83.013419} {\bibfield  {journal} {\bibinfo  {journal} {Phys.
  Rev. A}\ }\textbf {\bibinfo {volume} {83}},\ \bibinfo {pages} {013419}
  (\bibinfo {year} {2011})}\BibitemShut {NoStop}%
\bibitem [{\citenamefont {Baggesen}\ \emph {et~al.}(2012)\citenamefont
  {Baggesen}, \citenamefont {Lindroth},\ and\ \citenamefont
  {Madsen}}]{baggesen2012}%
  \BibitemOpen
  \bibfield  {author} {\bibinfo {author} {\bibfnamefont {J.~C.}\ \bibnamefont
  {Baggesen}}, \bibinfo {author} {\bibfnamefont {E.}~\bibnamefont {Lindroth}},
  \ and\ \bibinfo {author} {\bibfnamefont {L.~B.}\ \bibnamefont {Madsen}},\
  }\bibfield  {title} {\enquote {\bibinfo {title} {Theory of attosecond
  absorption spectroscopy in krypton},}\ }\href {\doibase
  10.1103/PhysRevA.85.013415} {\bibfield  {journal} {\bibinfo  {journal} {Phys.
  Rev. A}\ }\textbf {\bibinfo {volume} {85}},\ \bibinfo {pages} {013415}
  (\bibinfo {year} {2012})}\BibitemShut {NoStop}%
\bibitem [{\citenamefont {Yue}\ and\ \citenamefont {Madsen}(2013)}]{yue2013}%
  \BibitemOpen
  \bibfield  {author} {\bibinfo {author} {\bibfnamefont {L.}~\bibnamefont
  {Yue}}\ and\ \bibinfo {author} {\bibfnamefont {L.~B.}\ \bibnamefont
  {Madsen}},\ }\bibfield  {title} {\enquote {\bibinfo {title} {Dissociation and
  dissociative ionization of h${}_{2}{}^{+}$ using the time-dependent surface
  flux method},}\ }\href {\doibase 10.1103/PhysRevA.88.063420} {\bibfield
  {journal} {\bibinfo  {journal} {Phys. Rev. A}\ }\textbf {\bibinfo {volume}
  {88}},\ \bibinfo {pages} {063420} (\bibinfo {year} {2013})}\BibitemShut
  {NoStop}%
\bibitem [{\citenamefont {Werner}\ and\ \citenamefont
  {Meyer}(1981)}]{werner1981}%
  \BibitemOpen
  \bibfield  {author} {\bibinfo {author} {\bibfnamefont {H.~J.}\ \bibnamefont
  {Werner}}\ and\ \bibinfo {author} {\bibfnamefont {W.}~\bibnamefont {Meyer}},\
  }\bibfield  {title} {\enquote {\bibinfo {title} {MCSCF study of the avoided
  curve crossing of the two lowest ${}^1\Sigma^+$ states of LiF},}\ }\href {\doibase
  10.1063/1.440893} {\bibfield  {journal} {\bibinfo  {journal} {The Journal of
  Chemical Physics}\ }\textbf {\bibinfo {volume} {74}},\ \bibinfo {pages}
  {5802--5807} (\bibinfo {year} {1981})},\ \Eprint
  {http://arxiv.org/abs/https://doi.org/10.1063/1.440893}
  {https://doi.org/10.1063/1.440893} \BibitemShut {NoStop}%
\bibitem [{\citenamefont {Stapelfeldt}\ and\ \citenamefont
  {Seideman}(2003)}]{stapelfeldt2003}%
  \BibitemOpen
  \bibfield  {author} {\bibinfo {author} {\bibfnamefont {H.}~\bibnamefont
  {Stapelfeldt}}\ and\ \bibinfo {author} {\bibfnamefont {T.}~\bibnamefont
  {Seideman}},\ }\bibfield  {title} {\enquote {\bibinfo {title} {Colloquium:
  Aligning molecules with strong laser pulses},}\ }\href {\doibase
  10.1103/RevModPhys.75.543} {\bibfield  {journal} {\bibinfo  {journal} {Rev.
  Mod. Phys.}\ }\textbf {\bibinfo {volume} {75}},\ \bibinfo {pages} {543--557}
  (\bibinfo {year} {2003})}\BibitemShut {NoStop}%
\bibitem [{\citenamefont {Holmegaard}\ \emph {et~al.}(2009)\citenamefont
  {Holmegaard}, \citenamefont {Nielsen}, \citenamefont {Nevo}, \citenamefont
  {Stapelfeldt}, \citenamefont {Filsinger}, \citenamefont {K\"upper},\ and\
  \citenamefont {Meijer}}]{holmegaard2009}%
  \BibitemOpen
  \bibfield  {author} {\bibinfo {author} {\bibfnamefont {L.}~\bibnamefont
  {Holmegaard}}, \bibinfo {author} {\bibfnamefont {J.~H.}\ \bibnamefont
  {Nielsen}}, \bibinfo {author} {\bibfnamefont {I.}~\bibnamefont {Nevo}},
  \bibinfo {author} {\bibfnamefont {H.}~\bibnamefont {Stapelfeldt}}, \bibinfo
  {author} {\bibfnamefont {F.}~\bibnamefont {Filsinger}}, \bibinfo {author}
  {\bibfnamefont {J.}~\bibnamefont {K\"upper}}, \ and\ \bibinfo {author}
  {\bibfnamefont {G.}~\bibnamefont {Meijer}},\ }\bibfield  {title} {\enquote
  {\bibinfo {title} {Laser-induced alignment and orientation of
  quantum-state-selected large molecules},}\ }\href {\doibase
  10.1103/PhysRevLett.102.023001} {\bibfield  {journal} {\bibinfo  {journal}
  {Phys. Rev. Lett.}\ }\textbf {\bibinfo {volume} {102}},\ \bibinfo {pages}
  {023001} (\bibinfo {year} {2009})}\BibitemShut {NoStop}%
\bibitem [{\citenamefont {De}\ \emph {et~al.}(2009)\citenamefont {De},
  \citenamefont {Znakovskaya}, \citenamefont {Ray}, \citenamefont {Anis},
  \citenamefont {Johnson}, \citenamefont {Bocharova}, \citenamefont
  {Magrakvelidze}, \citenamefont {Esry}, \citenamefont {Cocke}, \citenamefont
  {Litvinyuk},\ and\ \citenamefont {Kling}}]{de2009}%
  \BibitemOpen
  \bibfield  {author} {\bibinfo {author} {\bibfnamefont {S.}~\bibnamefont
  {De}}, \bibinfo {author} {\bibfnamefont {I.}~\bibnamefont {Znakovskaya}},
  \bibinfo {author} {\bibfnamefont {D.}~\bibnamefont {Ray}}, \bibinfo {author}
  {\bibfnamefont {F.}~\bibnamefont {Anis}}, \bibinfo {author} {\bibfnamefont
  {N.~G.}\ \bibnamefont {Johnson}}, \bibinfo {author} {\bibfnamefont {I.~A.}\
  \bibnamefont {Bocharova}}, \bibinfo {author} {\bibfnamefont {M.}~\bibnamefont
  {Magrakvelidze}}, \bibinfo {author} {\bibfnamefont {B.~D.}\ \bibnamefont
  {Esry}}, \bibinfo {author} {\bibfnamefont {C.~L.}\ \bibnamefont {Cocke}},
  \bibinfo {author} {\bibfnamefont {I.~V.}\ \bibnamefont {Litvinyuk}}, \ and\
  \bibinfo {author} {\bibfnamefont {M.~F.}\ \bibnamefont {Kling}},\ }\bibfield
  {title} {\enquote {\bibinfo {title} {Field-free orientation of co molecules
  by femtosecond two-color laser fields},}\ }\href {\doibase
  10.1103/PhysRevLett.103.153002} {\bibfield  {journal} {\bibinfo  {journal}
  {Phys. Rev. Lett.}\ }\textbf {\bibinfo {volume} {103}},\ \bibinfo {pages}
  {153002} (\bibinfo {year} {2009})}\BibitemShut {NoStop}%
\bibitem [{\citenamefont {Oda}\ \emph {et~al.}(2010)\citenamefont {Oda},
  \citenamefont {Hita}, \citenamefont {Minemoto},\ and\ \citenamefont
  {Sakai}}]{oda2010}%
  \BibitemOpen
  \bibfield  {author} {\bibinfo {author} {\bibfnamefont {K.}~\bibnamefont
  {Oda}}, \bibinfo {author} {\bibfnamefont {M.}~\bibnamefont {Hita}}, \bibinfo
  {author} {\bibfnamefont {S.}~\bibnamefont {Minemoto}}, \ and\ \bibinfo
  {author} {\bibfnamefont {H.}~\bibnamefont {Sakai}},\ }\bibfield  {title}
  {\enquote {\bibinfo {title} {All-optical molecular orientation},}\ }\href
  {\doibase 10.1103/PhysRevLett.104.213901} {\bibfield  {journal} {\bibinfo
  {journal} {Phys. Rev. Lett.}\ }\textbf {\bibinfo {volume} {104}},\ \bibinfo
  {pages} {213901} (\bibinfo {year} {2010})}\BibitemShut {NoStop}%
\bibitem [{\citenamefont {Frumker}\ \emph {et~al.}(2012)\citenamefont
  {Frumker}, \citenamefont {Hebeisen}, \citenamefont {Kajumba}, \citenamefont
  {Bertrand}, \citenamefont {W\"orner}, \citenamefont {Spanner}, \citenamefont
  {Villeneuve}, \citenamefont {Naumov},\ and\ \citenamefont
  {Corkum}}]{frumker2012}%
  \BibitemOpen
  \bibfield  {author} {\bibinfo {author} {\bibfnamefont {E.}~\bibnamefont
  {Frumker}}, \bibinfo {author} {\bibfnamefont {C.~T.}\ \bibnamefont
  {Hebeisen}}, \bibinfo {author} {\bibfnamefont {N.}~\bibnamefont {Kajumba}},
  \bibinfo {author} {\bibfnamefont {J.~B.}\ \bibnamefont {Bertrand}}, \bibinfo
  {author} {\bibfnamefont {H.~J.}\ \bibnamefont {W\"orner}}, \bibinfo {author}
  {\bibfnamefont {M.}~\bibnamefont {Spanner}}, \bibinfo {author} {\bibfnamefont
  {D.~M.}\ \bibnamefont {Villeneuve}}, \bibinfo {author} {\bibfnamefont
  {A.}~\bibnamefont {Naumov}}, \ and\ \bibinfo {author} {\bibfnamefont {P.~B.}\
  \bibnamefont {Corkum}},\ }\bibfield  {title} {\enquote {\bibinfo {title}
  {Oriented rotational wave-packet dynamics studies via high harmonic
  generation},}\ }\href {\doibase 10.1103/PhysRevLett.109.113901} {\bibfield
  {journal} {\bibinfo  {journal} {Phys. Rev. Lett.}\ }\textbf {\bibinfo
  {volume} {109}},\ \bibinfo {pages} {113901} (\bibinfo {year}
  {2012})}\BibitemShut {NoStop}%
\bibitem [{\citenamefont {Kammler}(2007)}]{kammler2007}%
  \BibitemOpen
  \bibfield  {author} {\bibinfo {author} {\bibfnamefont {D.~W.}\ \bibnamefont
  {Kammler}},\ }\href@noop {} {\emph {\bibinfo {title} {A first course in
  Fourier analysis}}}\ (\bibinfo  {publisher} {Cambridge University Press,
  Cambridge},\ \bibinfo {year} {2007})\BibitemShut {NoStop}%
\end{thebibliography}
\end{document}